\documentclass[twocolumn]{aastex63}

\usepackage{xspace}

\newcommand{\jetset}{\texttt{JetSeT}\xspace}

\usepackage[colorinlistoftodos]{todonotes}
\usepackage{soul}
\usepackage{cancel}
\usepackage{ulem}
\usepackage{graphicx}  

\received{-}
\revised{-}
\accepted{-}
\submitjournal{ApJ}

\shorttitle{Broadband View of MAXI J\(1820+070\)}
\shortauthors{Rodi et al.}
\graphicspath{{./}{figures/}}
\begin{document}

\title{A broadband view on microquasar MAXI J\(1820+070\) during the 2018 outburst}

%%%%%%%%%%%%%%%%%%%%%%%%%%%%%%%%%%%%%%%%%%%%%%%%%%%%%%%%%%%%%%%%%%%%%%%%%%%%%%%%
%%%%%%%%%%%%%%%%%%%%%%%%%%%%%%%%%%%%%%%%%%%%%%%%%%%%%%%%%%%%%%%%%%%%%%%%%%%%%%%%

\correspondingauthor{J. Rodi}
\email{james.rodi@inaf.it}

\author[0000-0003-2126-5908]{J. Rodi}
\affiliation{INAF - Istituto di Astrofisica e Planetologia Spaziali, via Fosso del Cavaliere 100, 00133 Roma, Italy}

\author[0000-0002-8186-3793]{A. Tramacere}
\affiliation{Department of Astronomy, University of Geneva,
              Ch. d'Ecogia 16, 1290, Versoix, Switzerland}

\author[0000-0001-6286-1744]{F. Onori}
\affiliation{INAF - Istituto di Astrofisica e Planetologia Spaziali, via Fosso del Cavaliere 100, 00133 Roma, Italy}
\affiliation{INAF - Osservatorio Astronomico d'Abruzzo, via M. Maggini snc, I-64100 Teramo, Italy}

\author[0000-0002-5182-6289]{G. Bruni}
\affiliation{INAF - Istituto di Astrofisica e Planetologia Spaziali, via Fosso del Cavaliere 100, 00133 Roma, Italy}

\author{C. Sánchez-Fernández}
\affiliation{European Space Astronomy Centre (ESA/ESAC), Science Operations Department, 28691 Villanueva dela Cañada, Madrid, Spain}

\author{M. Fiocchi}
\affiliation{INAF - Istituto di Astrofisica e Planetologia Spaziali, via Fosso del Cavaliere 100, 00133 Roma, Italy}

\author{L. Natalucci}
\affiliation{INAF - Istituto di Astrofisica e Planetologia Spaziali, via Fosso del Cavaliere 100, 00133 Roma, Italy}

\author{P. Ubertini}
\affiliation{INAF - Istituto di Astrofisica e Planetologia Spaziali, via Fosso del Cavaliere 100, 00133 Roma, Italy}

%
%\collaboration{1}{(AAS Journals Data Scientists collaboration)}
%
%\author{Butler Burton}
%\affiliation{Leiden University}
%\affiliation{AAS Journals Associate Editor-in-Chief}
%\nocollaboration{1}
%\altaffiliation{AASTeX v6+ programmer}
%\collaboration{1}{(LaTeX collaboration)}
%\affiliation{AAS Director of Publishing}
%\affiliation{American Astronomical Society \\
%1667 K Street NW, Suite 800 \\
%\author{Scott Chernoff}
%\affiliation{IOP Publishing, Washington, DC 20005}




%%%%%%%%%%%%%%%%%%%%%%%%%%%%%%%%%%%%%%%%%%%%%%%%%%%%%%%%%%%%%%%%%%%%%%%%%%%%%%%%
%%%%%%%%%%%%%%%%%%%%%%%%%%%%%%%%%%%%%%%%%%%%%%%%%%%%%%%%%%%%%%%%%%%%%%%%%%%%%%%%

\begin{abstract}

The microquasar MAXI J\(1820+070\) went into outburst from mid-March until mid-July 2018 with several faint rebrightenings afterwards.  With a peak flux of approximately 4 Crab in the \(20-50\) keV, energy range the source was monitored across the electromagnetic spectrum with detections from radio to hard X-ray frequencies.  Using these multi-wavelength observations, we analyzed quasi-simultaneous observations from 12 April, near the peak of the outburst (\(\sim 23\) March).  Spectral analysis of the hard X-rays found a \(kT_e \sim 30 \) keV and \( \tau \sim 2\) with a \texttt{CompTT} model, indicative of an accreting black hole binary in the hard state.  The flat/inverted radio spectrum and the accretion disk winds seen at optical wavelengths are also consistent with the hard state.  Then we constructed a spectral energy distribution spanning \(\sim 12\) orders of magnitude using modelling in \texttt{JetSeT}.  The model is composed of an irradiated disk with a Compton hump and a leptonic jet with an acceleration region and a synchrotron-dominated cooling region.  \texttt{JetSeT} finds the spectrum is dominated by jet emission up to approximately \(10^{14}\) Hz after which disk and coronal emission dominate.  The acceleration region has a magnetic field of \( B \sim 1.6 \times 10^4 \) G, a cross section of \(R \sim 2.8 \times 10^{9} \) cm, and a flat radio spectral shape naturally obtained from the synchroton cooling of the accelerated electrons.  The jet luminosity of \(> 8 \times 10^{37} \) erg/s (\(> 0.15L_{Edd}\)) compared to an accretion luminosity of \( \sim 6 \times 10^{37}\) erg/s, assuming a distance of 3 kpc.  Because these two values are comparable, it is possible the jet is powered predominately via accretion with only a small contribution needed from the Blanford-Znajek mechanism from the reportedly slowly spinning black hole.

\end{abstract}



%%%%%%%%%%%%%%%%%%%%%%%%%%%%%%%%%%%%%%%%%%%%%%%%%%%%%%%%%%%%%%%%%%%%%%%%%%%%%%%%
%%%%%%%%%%%%%%%%%%%%%%%%%%%%%%%%%%%%%%%%%%%%%%%%%%%%%%%%%%%%%%%%%%%%%%%%%%%%%%%%

\section{Introduction} \label{sec:intro}
%Microquasars are cool. \cite{2020MNRAS.492.3657F} [...]

%The black hole X-ray binary MAXI J1820+070 was discovered in March 2018 (...). This prompted a multi-wavelength campaign involving several instruments from radio to gamma-ray band.  \cite{2020NatAs.tmp....2B, 2020MNRAS.493L..81A} [...]

The term "microquasar" was first applied to the persistent black hole candidate (BHC) 1E \(1740.7-2942\) after detecting radio jets from the known hard X-ray source \citep{1992Natur.358..215M} that were similar to radio-loud active galactic nuclei (AGNs).  Jets were later found to be common features in accreting BH systems in the hard state.  Multi-wavelength studies showed correlations between radio and X-ray luminosities \citep{2003MNRAS.344...60G}, indicating a relationship between the emission mechanisms despite a large physical separation between the two.  Additionally, this correlation holds also for supermassive BHs in AGN  when accounting for mass \citep{2003MNRAS.345.1057M}, thus linking the mechanisms in stellar mass and supermassive BHs.  Therefore understanding the jet and X-ray components in microquasars can shed light on AGN.  

The low-mass X-ray binary MAXI J\(1820+070\) (=ASASSN-18ey) was first detected on 6.59 March 2018\footnote{http://www.astronomy.ohio-state.edu/asassn/transients.html} with the All-Sky Automated Survey for SuperNovae \citep{2014ApJ...788...48S} and was detected \( \sim 6\) days later by the \textit{MAXI}/GSC at 11 March 2018 19:48 UTC \citep{2018ATel11399....1K}.  With a peak flux of \(\sim 4\) Crab in the \(20-50\) keV energy band \citep{2019ApJ...870...92R} and a long decay, the source was a good candidate for numerous observing campaigns across the electromagnetic (EM) spectrum (e.g. \citealt{2019ApJ...879L...4M,2018ApJ...867L...9T,2020ApJ...889..142S,2020NatAs.tmp....2B}) to explore various aspects of the source.  Combining observations from various campaigns enables studying the various emission processes together.  

Therefore we compiled quasi-simultaneous observations from public archives, Astronomer's Telegrams, and Gamma-ray Coordination Network Circulars to construct the widest possible frequency range, we were able to find detections covering nearly 12 orders of magnitude from the meter-wavelength frequencies to hard X-rays on 12 April 2018 (MJD 58220).  With this spectral energy distribution (SED), we studied the spectral components independently  before investigating them jointly by constructing a model consisting of a leptonic jet, an irradiated disk, and a corona, using the \texttt{JetSeT} software\footnote{\url{https://jetset.readthedocs.io/en/latest/}}.

\begin{figure}
    \centering
    \includegraphics[angle=180,scale=0.5,trim = 20mm 30mm 43mm 10mm]{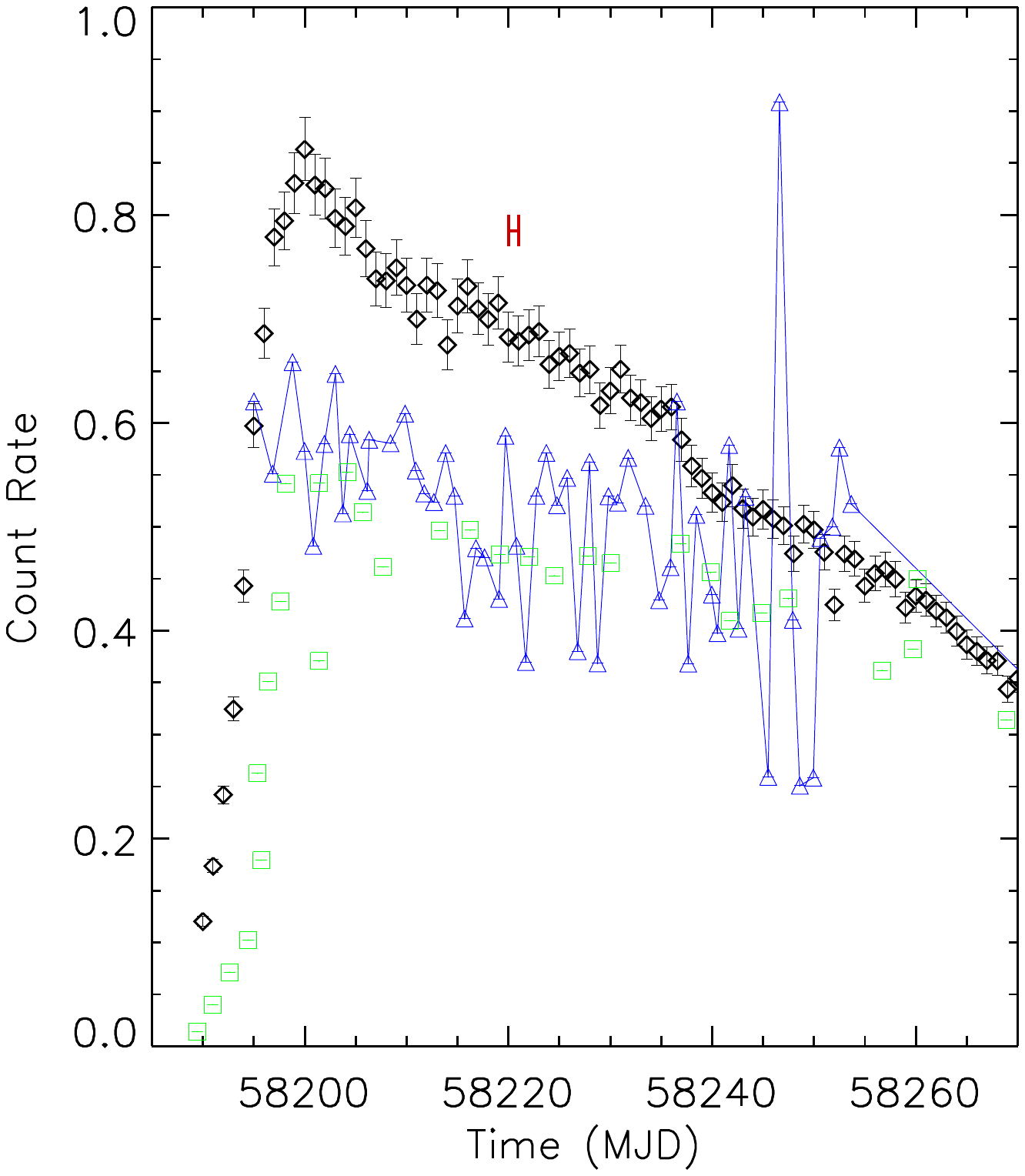}
    \caption{(Top) The light curves for \textit{Swift}/BAT \(15-50\) keV (black diamonds), \(0.3-10\) keV \textit{Swift}/XRT (green squares), and 4.7 GHz RATAN (blue triangles) for the initial phase of the outburst.  The time span of  observations analyzed in this work are denoted in red.}
    \label{fig:bat_lc}
\end{figure}
%%%%%%%%%%%%%%%%%%%%%%%%%%%%%%%%%%%%%%%%%%%%%%%%%%%%%%%%%%%%%%%%%%%%%%%%%%%%%%%%
%%%%%%%%%%%%%%%%%%%%%%%%%%%%%%%%%%%%%%%%%%%%%%%%%%%%%%%%%%%%%%%%%%%%%%%%%%%%%%%%

\section{Observations}
Figure~\ref{fig:bat_lc} shows the initial period of MAXI J\(1820+070\) outburst in several wavelengths across the spectrum using data \textit{Swift}/BAT (black diamonds), \textit{Swift}/XRT (green squares) \citep{2020ApJ...889..142S}, and 4.7 GHz RATAN (blue triangles) \citep{2018ATel11539....1T}.  The XRT and RATAN data have been normalized to be on the same scale of the BAT data.  The period of the observations used in this work are bracketed in red.

In the following, we give information about the different simultaneous observations collected from archives, covering different bands from radio to gamma-ray on April 12, 2018. Further details can be found in Table~\ref{tab:obs}.

\subsection{JVLA}
We retrieved calibrated Karl G. Jansky Very Large Array (VLA) data for experiment VLA/18A-470 from the National Radio Astronomy (NRAO) online archive.  The JVLA antennas, in A configuration, were split in 3 subarrays in order to obtain simultaneous observations at 6 different central frequencies (4.7 GHz, 7.5 GHz, 8.5 GHz, 11 GHz, 20.7 GHz, 25.5 GHz). Data were imaged using {\tt{CASA}} ({\tt{Common Astronomy Software Applications}} package) version 5.6.2\footnote{\hyperref[https://casa.nrao.edu]{https://casa.nrao.edu}} following standard procedures. 

\subsection{ALMA}
Atacama Large Millimiter/submillimiter Array (ALMA) data for project 2017.1.01103.T were retrieved from the ESO archive, and pipelined at the the Italian node of the European ALMA Regional Centre (INAF-Istituto di Radioastronomia, Bologna). Imaging was 
performed with {\tt{CASA}} version 5.1.1, separately for each one of the 4 spectral windows (spw) present in the data (Band 7, spw 5, 7, 9, 11) corresponding to the following central frequencies: 336 GHz, 338 GHz, 348 GHz, 350 GHz \citep{2018MNRAS.478.1512B}.

\subsection{VLT/X-shooter}
A number of observations of MAXI J\(1820+070\) were performed with the X-shooter spectrograph \cite[][]{vernet11} in the framework of the ESO program 0101.D-0356(A). We retrieved the processed spectra obtained during the 2018 outburst on 12 April from the European Southern Observatory (ESO) archive science portal. These data have been reduced by using the ESO X-shooter pipeline
V2.7.0 and cover the 3000-25000 \AA wavelength range. The observations were conduced in nodding configuration with the slit oriented at the parallactic angle and using slit widths of 1.3$\arcsec\times$11, 1.2$\arcsec\times$11 and 1.2$\arcsec\times$11 for the UVB, VIS and NIR arm, respectively. This configuration yields a spectral resolution R=$\lambda$/$\Delta \lambda$ of 4100, 6500 and 4300 for the UVB, VIS and NIR arm, respectively. The observing conditions where good with a seeing of 0.47$\arcsec$ and an average airmass of the source during the acquisition of 1.3$\arcsec$. The total exposure times are 1640 s, 1300 s and 1520 s for the UVB, VIS, and NIR arm, respectively. %The acquisition image was taken using the $i^\prime$ filter.\\
The reduced spectra have been corrected for the foreground extinction using the Cardelli function \cite[]{cardelli89} with R(V)=3.1 and A$_V$=0.627 mag \cite[][via the NASA/IPAC Extragalactic Database (NED)]{Schlafly11}.

In order to estimate the slit loss effect in the X-shooter spectra, we first applied standard aperture photometry on the $i^\prime$ acquisition image using the \texttt{iraf} task \texttt{phot}. The zero point was calibrated using the stars in the Panoramic Survey Telescope and Rapid Response System \cite[Pan-STARRS1][]{Flewelling2016} catalog. 

%The magnitude uncertainty is calculated using the photometric error from the aperture photometry procedure and the standard deviation of the catalog sources used to derive the zero point. 
%We take the foreground (Milky Way) extinction towards MAXI to be $A_V$=.. mag \cite[][via the NASA/IPAC Extragalactic Database (NED)]{Schlafly11}. All the input magnitudes have been corrected for the Galactic extinction from \cite{Schlafly11} which assume a reddening law with Rv = 3.1
From the aperture photometry we obtain an $i^\prime$ band apparent magnitude of m$_{AB}$= (12.20 $\pm$ 0.11) mag, corrected for foreground extinction. The derived flux at the filter central wavelength is $\lambda$F$_{i}$ = (2.01$\pm$0.2)$\times$10$^{-10}$ erg s$^{-1}$ cm$^{-2}$ which is in agreement with the average flux measured from the spectrum in the 7300-7600 \AA\/ wavelength range: $\lambda$F$_{i}$=(1.8$\pm$0.7)$\times$10$^{-10}$ erg s$^{-1}$ cm$^{-2}$. %calculated from statistics in a region between 730-760 nm with qfitsview. Used average vales and standard dev. Lambda central is 7500 \AA. 

%We derive the optical fluxes in the $i^\prime$ and $z^\prime$ band by using the X-shooter VIS spectrum, while the Johnson \textit{V} band values have been obtained from the \textit{INTEGRAL}/OMC data. 

%derive spectral resolution.

\subsection{\textit{XMM-Newton}/EPIC-pn}
\textit{XMM-Newton} ToO observations were carried out from 2018-04-12 07:27:58 to 09:39:28 UTC (obsid 0820880501) using burst mode.  The European Photon Imaging Camera (EPIC)-pn data were analyzed using the standard procedures with the Science Analysis System (SAS) software version \texttt{xmmsas\_20190531\_1155-18.0.0}\footnote{https://www.cosmos.esa.int/web/xmm-newton/sas-threads}.

\subsection{INTEGRAL} 
The \textit{INTErnational Gamma-Ray Astrophysics Laboratory} (\textit{INTEGRAL}) observed MAXI J\(1820+070\) every 2--3 days between March 16, and May 8,  via a series of Target of Opportunity (ToO) observations. For this work, we selected the data covering the interval (UTC) 11 April 2018 23:41:01 to  12 April 2018 11 14:00:21 (\textit{INTEGRAL} revolution 1941).
Here we focus on the analysis of data provided by the  Integral Soft Gamma-Ray Imager (ISGRI; \(18-1000\) keV) placed on the upper layer of the detector plane of the Imager on Board the INTEGRAL Satellite (IBIS) telescope  (\citealt{Ubertini2003}) and by the Optical Monitoring Camera (OMC) (\(500-600\) nm) instruments.  The data were analyzed using the  Offline  Science  Analysis software (OSA) v11.0 available at the \textit{INTEGRAL} Science Data Center (ISDC).\footnote{\hyperref[https://www.isdc.unige.ch/integral/analysis]{https://www.isdc.unige.ch/integral/analysis}} We followed standard analysis procedures.

%As part of a set of Target of Opportunity (ToO) observations, the \textit{INTErnational Gamma-Ray Astrophysics Laboratory} (\textit{INTEGRAL}) observed MAXI J\(1820+070\) for nearly two months.  
%In this work we analyze data from the IBIS/ISGRI (\(18-1000\) keV; \citealt{Ubertini2003}), JEM-X (\(3-35\) keV), and the Optical Monitoring Camera (OMC) instruments.  We selected data covering 23:41:01 UTC 11 April 2018 to 14:00:21 12 April 2018 (\textit{INTEGRAL} revolution 1941). These data were analyzed using the  \textcolor{purple}{Offline  Science  Analysis software} (OSA) v11.0 available at the \textit{INTEGRAL} Science Data Center (ISDC)\footnote{\hyperref[https://www.isdc.unige.ch/integral/analysis]{https://www.isdc.unige.ch/integral/analysis}} \textcolor{purple}{following standard analysis procedures}.

%%%%%%%%%%%%%%%%%%%%%%%%%%%%%%%%%%%%%%%%%%%%%%%%%%%%%%%%%%%%%%%%%%%%%%%%%%%%%%%%
\begin{deluxetable}{ccccc}
\tablenum{1}
\tablecaption{MAXI J\(1820+070\) observations log \label{tab:obs}}
\tablewidth{0pt}
\tablehead{
\colhead{Instrument} & 
%\colhead{Energy band} &
\colhead{Start Time (UTC)} &
\colhead{Stop Time (UTC)} \\ 
%\colhead{Exposure Time (s)} \\
%\multicolumn2c{(kpc)} & \colhead{Constellation} & \colhead{(mag)}
}
%\decimalcolnumbers
\startdata
VLITE          & 07:25:00 12-04-2018  & 13:07:00 12-04-2018. \\
JVLA           & 07:15:00 12-04-2018  & 13:14:50 12-04-2018  \\
ALMA           & 08:13:18 12-04-2018  & 09:20:16 12-04-2018  \\
X-shooter      & 07:41:08 12-04-2018  & 08:15:48 12-04-2018  \\ %& 1640; 1300; 1520\\ % exposure times for UVB, VIS and NIR, respectively
EPIC-pn        & 07:27:58 12-04-2018  & 09:39:28 12-04-2018  \\
OMC            & 23:41:01 11-04-2018  & 14:00:21 12-04-2018  \\
ISGRI          & 23:41:01 11-04-2018  & 14:00:21 12-04-2018  \\ %& \(2.416 \times 10^4\)  \\
\enddata
%\tablecomments{X-shooter exposure times are indicated for UVB, VIS and NIR arms, respectively}
\end{deluxetable}
%%%%%%%%%%%%%%%%%%%%%%%%%%%%%%%%%%%%%%%%%%%%%%%%%%%%%%%%%%%%%%%%%%%%%%%%%%%%%%%%

%%%%%%%%%%%%%%%%%%%%%%%%%%%%%%%%%%%%%%%%%%%%%%%%%%%%%%%%%%%%%%%%%%%%%%%%%%%%%%%%
\begin{deluxetable}{ccccc}
\tablenum{2}
\tablecaption{Collected flux densities for the jet modelling. \label{tab:fluxes}}
\tablewidth{0pt}
\tablehead{
\colhead{Instrument} & 
\colhead{Frequency} &
\colhead{Flux density} \\
\colhead{} & 
\colhead{(Hz)} &
\colhead{(Jy)}
%\multicolumn2c{(kpc)} & \colhead{Constellation} & \colhead{(mag)}
}
%\decimalcolnumbers
\startdata
VLITE       & 3.39E+08  & 0.033$\pm$5.3      \\
JVLA        & 4.70E+09 & 0.0469$\pm$0.0047  \\
            & 7.50E+09 & 0.0488$\pm$0.0029  \\
            & 8.50E+09 & 0.0479$\pm$0.0048  \\
            & 1.10E+10 & 0.0483$\pm$0.0048  \\
            & 2.07E+10 & 0.0525$\pm$0.0054  \\
            & 2.55E+10 & 0.0530$\pm$0.0054  \\
ALMA        & 3.36E+11 & 0.116$\pm$0.006    \\      
            & 3.38E+11 & 0.114$\pm$0.006    \\
            & 3.48E+11 & 0.115$\pm$0.006    \\
            & 3.50E+11 & 0.110$\pm$0.006    \\
X-shooter   & 1.37E+14 & 0.0920$\pm$0.0002  \\  
            & 1.81E+14 & 0.0757$\pm$0.0012  \\  
            & 2.40E+14 & 0.0652$\pm$0.0003  \\
            & 2.86E+14 & 0.0634$\pm$0.0006  \\
            & 3.79E+14 & 0.0689$\pm$0.0007  \\
            & 4.00E+14 & 0.0709$\pm$0.0002  \\
            & 4.42E+14 & 0.0743$\pm$0.0003  \\
            & 4.87E+14 & 0.0775$\pm$0.0005  \\
            & 5.15E+14 & 0.0794$\pm$0.0002  \\
            & 5.32E+14 & 0.0803$\pm$0.0003  \\
            & 5.76E+14 & 0.0831$\pm$0.0005  \\
            & 6.27E+14 & 0.0868$\pm$0.0003  \\
            & 7.00E+14 & 0.0952$\pm$0.0004  \\
OMC         & 5.66E+14 & 0.0813$\pm$0.0071  \\             
%Epic-pn     &   &    \\      
%JEM-X       &   &    \\    
%ISGRI       &   &    \\    
\enddata
%\tablecomments{}
\end{deluxetable}
%%%%%%%%%%%%%%%%%%%%%%%%%%%%%%%%%%%%%%%%%%%%%%%%%%%%%%%%%%%%%%%%%%%%%%%%%%%%%%%%

\begin{figure}
    \centering
    \includegraphics[width=8.5cm]{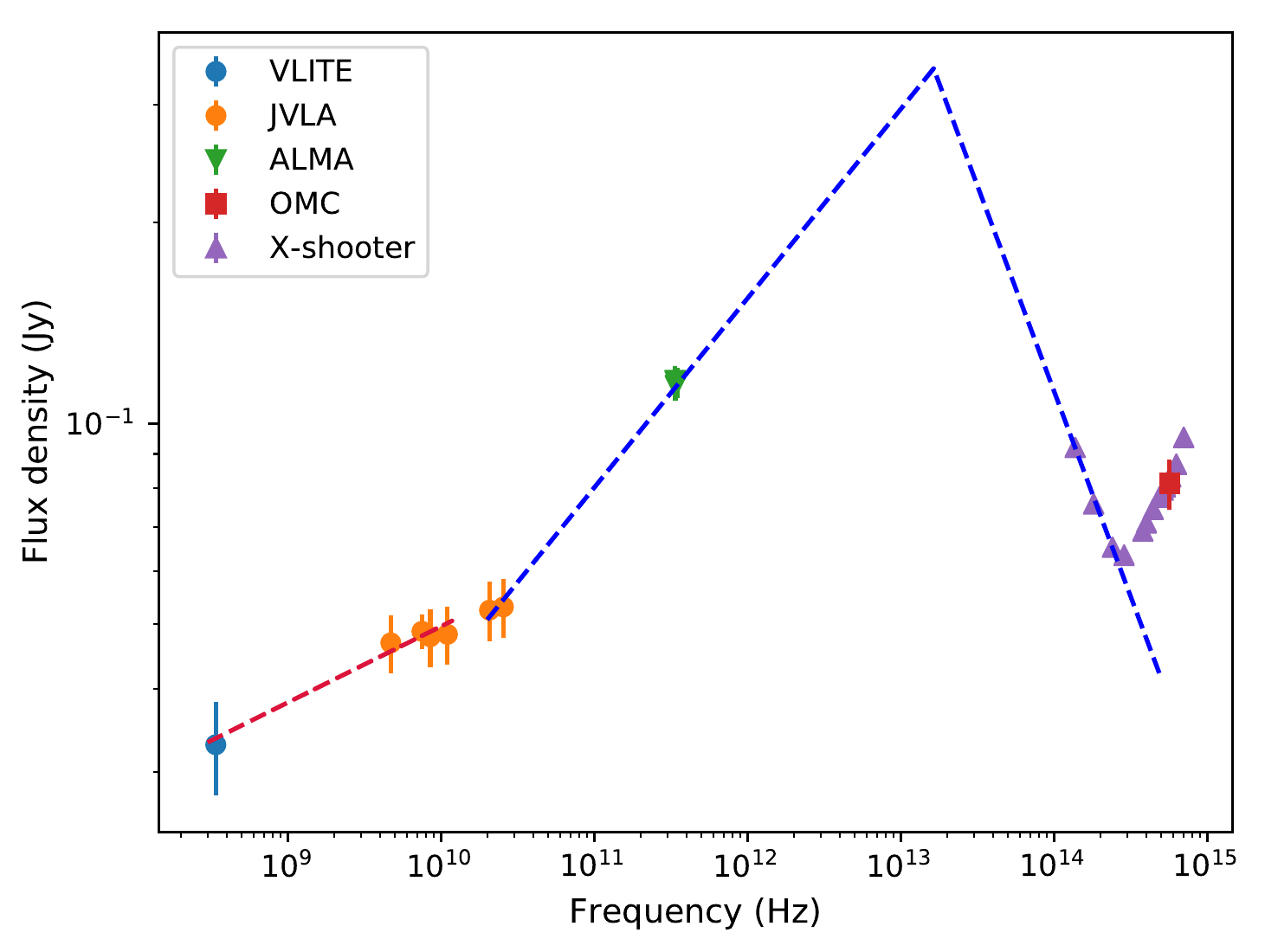}
    \caption{Jet emission between radio and optical bands, as reconstructed from VLITE, JVLA, ALMA, and X-shooter observations. The blue dashed line is a broken power law, used to identify the synchrotron peak frequency and flux density. The red line is a power-law fit of the most expanded region of the jet, considered as a physically separated component. The optical flux from OMC is also shown, although not considered for the fit.  }
    \label{fig:jet_compact}
\end{figure}

%%%%%%%%%%%%%%%%%%%%%%%%%%%%%%%%%%%%%%%%%%%%%%%%%%%%%%%%%%%%%%%%%%%%%%%%%%%%%%%%

\section{Results and discussion}

\subsection{The compact jet emission}

    With the collected flux densities between radio and UV bands, we built the jet SED. In addition to the JVLA and ALMA data mentioned above, we considered Very Large Array Low-band Ionosphere and Transient Experiment (VLITE, \citealt{2016SPIE.9906E..5BC}), also collected on 12 April 2018 \citep{2018ATel11540....1P}. We show in Figure~\ref{fig:jet_compact} data from JVLA, ALMA, X-shooter, and OMC. For X-shooter, we considered only the part of the spectrum not affected by absorption/emission features, and averaged values for each of these intervals to obtain continuum flux density values. In this way, we calculated 12 photometric measurements (Table~\ref{tab:fluxes}) covering the interval from NIR to UV. The overall shape of the X-shooter spectrum shows a break, resulting in a change of the spectral index, at about 2$\times$10$^{14}$ Hz. This is most probably the frequency at which the broadband SED is no longer dominated by the jet synchrotron emission, while the accretion disk thermal emission increases (see Sec.~\ref{Sec:SED}). The single value from OMC is in good agreement with the X-shooter photometry.

    %With these data, we adopted the method presented in previous works on microquasars in the literature (e.g. \citealt{2013MNRAS.429..815R}) in order to identify the synchrotron peak and the spectral slope of its optically thin and thick regions. We thus fitted the radio to optical SED with a broken power-law (the Astropy {\tt{BrokenPowerLaw1D}} function, see \citealt{astropy:2013,astropy:2018}) 
    
    We fitted the radio to optical data set with a broken-power law to identify the synchrotron peak frequency and the spectral slopes of the optically thin and thick regions \citep{2013MNRAS.429..815R}.  We used the Astropy {\tt{BrokenPowerLaw1D}} function (see \citealt{astropy:2013,astropy:2018}) adopting the {\tt{LevMarLSQFitter}} routine to perform a Levenberg-Marquardt least squares statistic. We excluded from the fit the X-shooter points above the third one (Optical and UV ranges) since they show a turn-up of the flux density liekly due to accretion disk emission. Similarly, we did not consider data points below 20 GHz, since they show a different slope, probably belonging to a physically distinct (and more expanded) radio jet component. We obtained a synchrotron peak frequency of 1.6$\pm$0.2$\times10^{13}$ Hz. The estimated slope for the optically thick part of the spectrum is $\alpha_{thick}=0.28\pm0.02$ and $\alpha_{thin}=-0.61\pm0.01$ for the optically thin one (adopting the convention $S\propto\nu^{\alpha}$, where $S$ is the flux density, $\nu$ the frequency, and $\alpha$ the spectral index). The slope of the lower frequency radio SED (0.3$-$10 GHz), fitted with the {\tt{PowerLaw1D}} Astropy function, is $\alpha=0.11\pm0.02$. In Sec.~\ref{Sec:SED}, a detailed physical model of this source is presented and provides a more precise estimate of the jet parameters.%{The value found for the synchrotron peak is in the range expected for the hard state ($\sim10^{11}-10^{14}$ , see e.g. \citealt{2013MNRAS.429..815R}).  
    
    %The slope of the lower frequency radio SED (0.3$-$10 GHz), fitted with the {\tt{PowerLaw1D}} Astropy function, is $\alpha=-0.11\pm0.02$. In Sec. \ref{Sec:SED}, a detailed physical model of this source is presented, able to provide a more precise estimate of the jet parameters.   

%%%%%%%%%%%%%%%%%%%%%%%%%%%%%%%%%%%%%%%%%%%%%%%%%%%%%%%%%%%%%%%%%%%%%%%%%%%%%%%%

\begin{figure*}
    \centering
    \includegraphics[width=\textwidth]{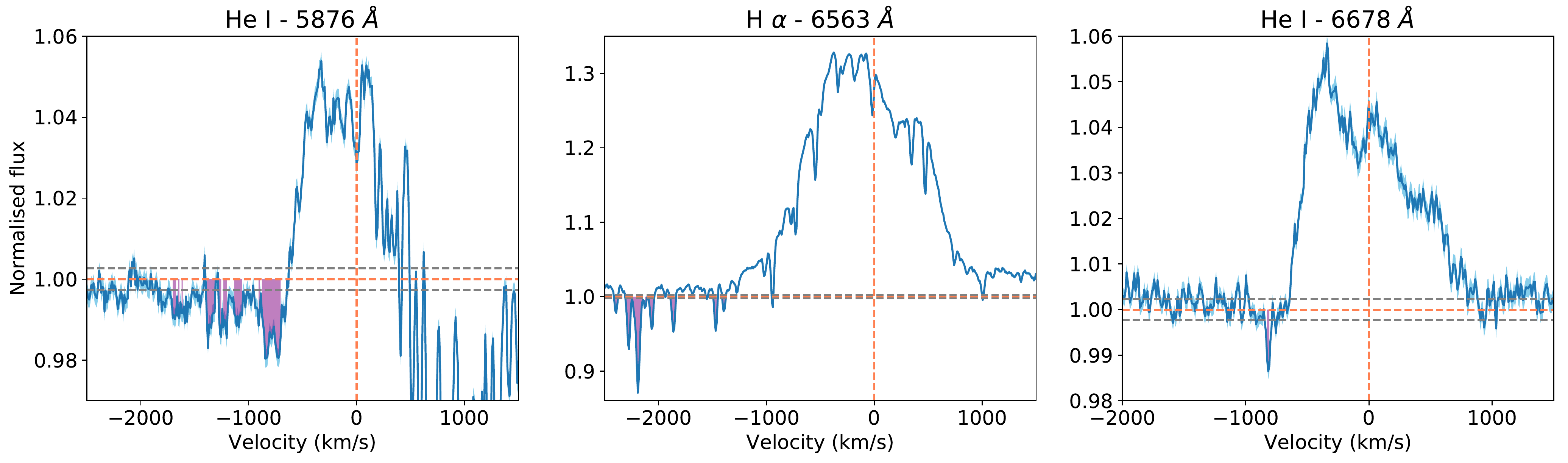}
    \caption{Accretion disk wind absorption features in the VLT/X-shooter optical spectrum. Normalised flux errors are shown as a cyan shaded area. Troughs deeper than 3$\times$RMS are highlighted in purple. The RMS value is 0.0027 for the He\, I $\lambda$5876 region, while 0.0023 for the H$\alpha$ and He\,I $\lambda$6678 regions. The grey dashed lines indicate the 1$\pm$RMS intervals.} %\textcolor{purple}{Looking at the He-I plot, I get the impression that continuum increases between -1000 and -2000kms, suggesting that also in this line there could be some degree of P-Cygni absorption (although at a lesser degree than in the HeI line). Have you checked this?.}}
    \label{fig:winds}
\end{figure*}

%%%%%%%%%%%%%%%%%%%%%%%%%%%%%%%%%%%%%%%%%%%%%%%%%%%%%%%%%%%%%%%%%%%%%%%%%%%%%%%%

\subsection{Accretion disk winds}
\cite{2019ApJ...879L...4M} discovered accretion disk winds in MAXI J\(1820+070\) during both the 2018 hard state rise and decay. Absorption wind signatures were detected in the blue wings of He\,I $\lambda$5876 and $\lambda$6678 emission lines, reaching a maximum terminal velocity ($v_t$) of 1200 km/s in one of the epochs. For H$\alpha$, both a blue-wing broadened emission line profile, implying a wind component of 1800 km/s, and a superimposed absorption trough with a $v_t$=1200 km/s were found. Those authors collected several epochs from 15 March to 4 November 2018, allowing them to follow the evolution of the winds from the hard state to the disappearance during soft state, and back. 

\begin{figure*}
   \includegraphics[width=8.8cm]{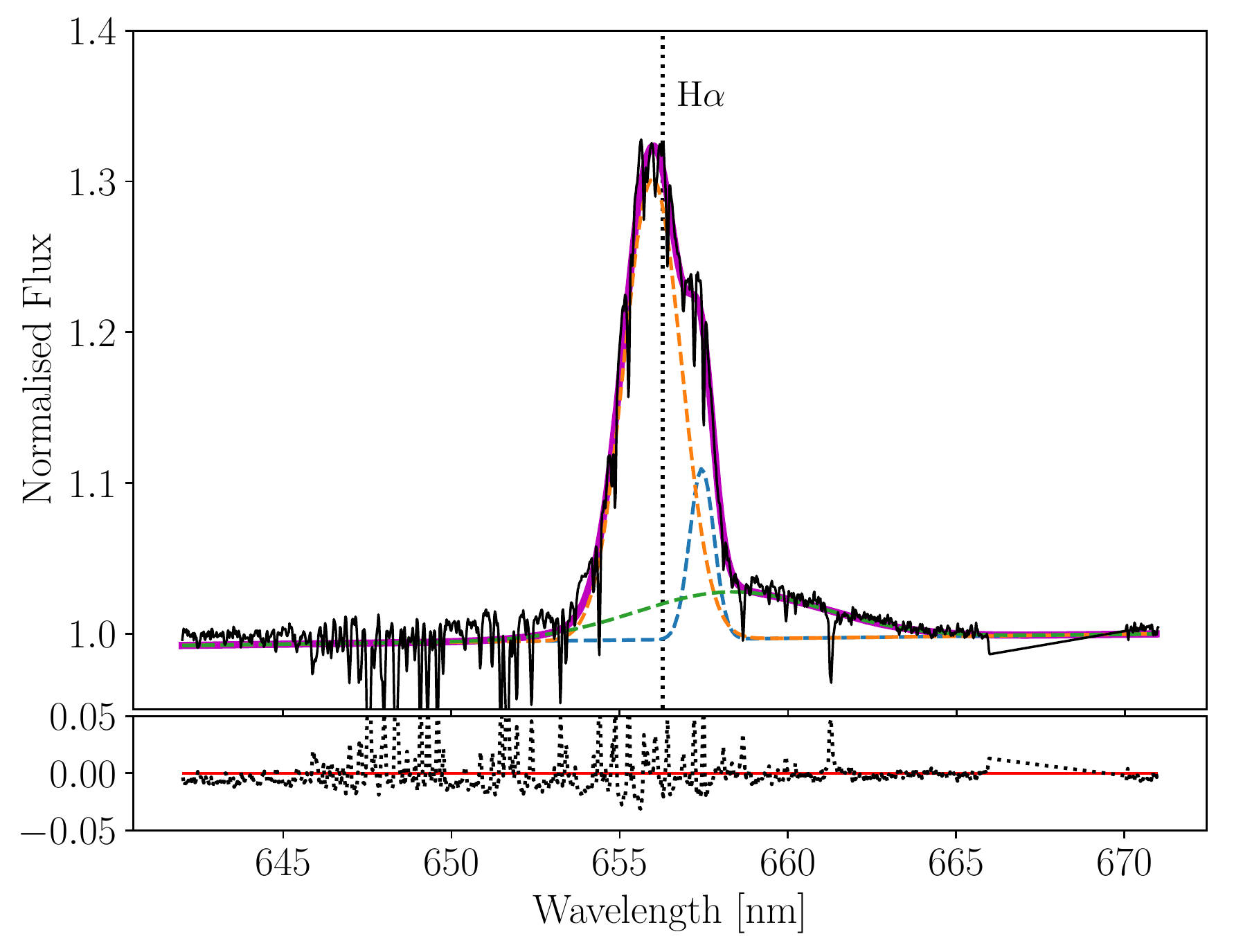}
   \includegraphics[width=8.8cm]{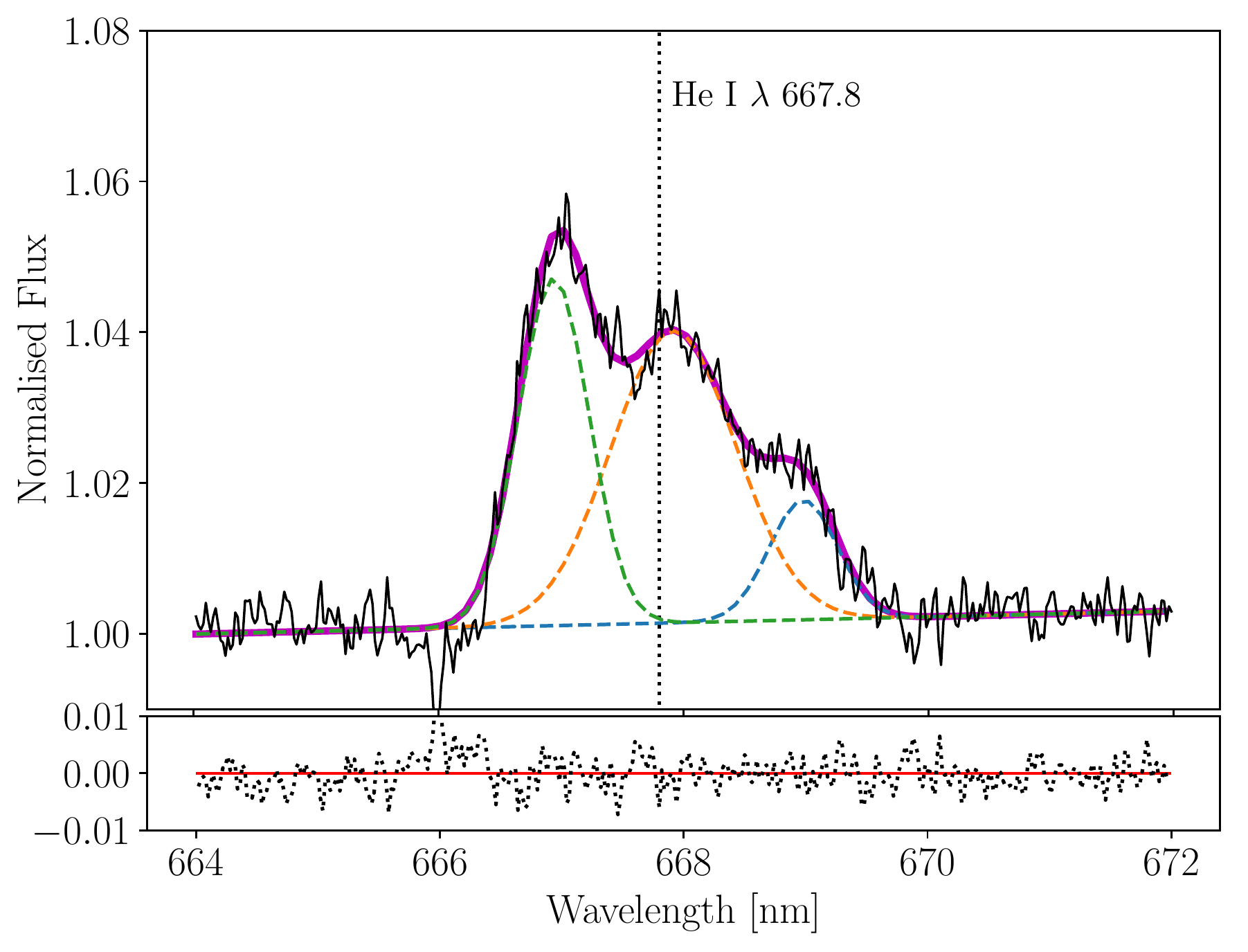}
   \caption{Fits of the emission line profiles for H$\alpha$ (left panel) and He\,I $\lambda$6678 (right panel). The residuals of the fits are shown at the bottom of each panel. The line profiles are modelled with Gaussian components (colored dashed lines). The total fitting model is represented by the magenta solid line.}
   \label{fig:linefit}
   \end{figure*}

The VLT/X-shooter data presented in this work add an epoch to monitoring in \cite{2019ApJ...879L...4M}, falling in an uncovered time window of one month between 26 March and 23 April. We normalized these spectra by dividing them for the continuum emission, fitted with a third order spline3 function by using the {\sc IRAF} task {\tt continuum}. These spectra are rich with emission lines from the UVB to the NIR arms. Following the \cite{2019ApJ...879L...4M} analysis, 
we explored the presence of wind signatures linked to the mentioned emission lines (He\,I and H$\alpha$), and found a significant absorption on the left wing of He\,I $\lambda$5876. In Fig. \ref{fig:winds} (left panel) we show the relative portion of the spectrum, with the absorption features highlighted in purple. We consider as bona-fide absorption troughs the ones with a dip of at least three times the continuum RMS. A prominent He\,I absorption feature is visible between -700 and -900 km/s, showing the same profile as the correspondent emission line. This one has a $v_t$ of 880 km/s. Further blue-ward absorption features are visible, but since they are narrower and not connected to the previous ones, we consider those as not related to the accretion disk wind. The same is true for the narrow absorption features detected blue-wards of the H$\alpha$ emission line (Fig. \ref{fig:winds}, central panel). For the He\,I $\lambda$6678 line, a single absorption trough is detected between -800 and -850 km/s (Fig. \ref{fig:winds}, right panel) with a $v_t$ of -825 km/s.\\ 
During this period, we observe strong asymmetries in the emission lines which are commonly observed in lines emitted from the disc, particularly in the He\,I $\lambda$ 6678 and in the H$\alpha$.
%During this period, we observe strong asymmetries in all the three emission lines, which are commonly observed in lines emitted from the disc. %In particular, the He\,I $\lambda$ 5876 shows a pronounced double-horned profile, while the He\,I $\lambda$ 6678 clearly shows the presence of three components. The H$\alpha$, instead, is characterized by two narrow components and a red, broad wing.
%especially in H$\alpha$ and He\,I $\lambda$6678.
Therefore, we explored line profile properties by applying multi-component Gaussian fits using the \texttt{python} packages \texttt{curvefit} and \texttt{leastsq}. 
In Figure \ref{fig:linefit} we show the result of this analysis for H$\alpha$ (left panel) and He\,I $\lambda$6678 (right panel). The H$\alpha$ line analysis has been performed in the wavelength region 644$-$670 nm, which includes the feature of interest and the local continuum. In this case we have ignored from the fit the He\,I emission line, which falls at the end of the analyzed wavelength range. The H$\alpha$ profile is well modelled by two narrow Gaussian components, and only one broad Gaussian component is needed to fit the red wing of the emission line. We note the absence of a blue-shifted broad wing, which has been observed in \cite{2019ApJ...879L...4M}, as well as p-cygni profiles signatures. However a forest of narrow absorption lines is clearly visible in the blue region of the H$\alpha$. %do we want to say something about this? could it be a signature of absorption material? 
The two narrow components are characterized by a central wavelength ($\lambda_{c}$) and a full width at half maximum (FWHM) of $\lambda_{c}$= (6574.4$\pm$0.4) \AA\, FWHM = (389$\pm$20) km/s and $\lambda_{c}$= (6559.8$\pm$ 0.3) \AA\, FWHM = (914$\pm$18) km/s, respectively. While the broad red wing is centered at $\lambda_{c}$= (6583.0$\pm$3.0) \AA\/ and has FWHM=(2982$\pm$210) km/s. From the redshift of the broad wing with respect to the H$\alpha$ rest frame wavelength we derive an outflow velocity of $v$= (923$\pm$14) km/s, while the separation between the two narrow components is $\sim$ 667 km/s. 

For the He\,I $\lambda$6678 line analysis we used the wavelength region \(664-672\) nm, which includes also the local continuum but excludes the wavelength range in which the H$\alpha$ falls. 
The He\,I $\lambda$6678 line profile is well modelled by three Gaussian components. The first one is well centered on the rest-frame He\,I wavelength with a $\lambda_{c}$= (6679.2 $\pm$ 0.4) \AA\ and has a FWHM=(536$\pm$54) km/s. The two remaining Gaussians are blueshifted and redshifted of $\sim$450 km/s with respect to the first component, and are characterized by $\lambda_{c}$= (6669.4 $\pm$ 0.2) \AA,  FWHM = (307$\pm$12) km/s and $\lambda_{c}$= (6689.7 $\pm$ 0.3) \AA,  FWHM = (295$\pm$29) km/s, respectively.

%\textcolor{purple}{I think most of the lines analysed by \cite{2019ApJ...879L...4M}, were double-peaked. It might be worth discussing why at this epoch the VLT/X-shooter data display also prominent emission at the rest-frame of the line.}%again, it could be interesting to discuss it, i.e. is the emission of the disk we are looking at? The different component intensities could they be due to disk inclinations?
%typical outflow tracers
%While the H$\alpha$ and He\,I $\lambda$5876 line profiles are similar to the one observed in \cite{2019ApJ...879L...4M} and can be explain as being at an intermediate phase of the source, the profile of He\,I $]\lambda$6678  
%\note{Add broad wing properties of the emission lines}.

%\textcolor{orange}{[FO]
As a whole, the detected optical disk wind features show properties with in between what was found in the hard and soft state epochs collected by \cite{2019ApJ...879L...4M}, confirming the decreasing trend of the optical wind between the two states of the source.

%As a whole, the detected wind features show intermediate properties with respect to the hard and soft state epochs collected by \cite{2019ApJ...879L...4M}, confirming the decreasing trend between the two states. \textcolor{green}{Munos-Darias says that optical winds are just during hard state so if we have optical winds then we should be in hard state. (JR)} 

%ionization properties from Bowen lines

%%%%%%%%%%%%%%%%%%%%%%%%%%%%%%%%%%%%%%%%%%%%%%%%%%%%%%%%%%%%%%%%%%%%%%%%%%%%%%%%

\subsection{Soft X-ray}
\label{Soft-X-ray-analysis}

%We fitted the XMM-Newton/EPIC-pn spectrum in the \(0.5-12\) keV energy range to study the soft X-ray emission from the system. Fitting the data with an absorbed power law model (\texttt{Tbabs*Powerlaw}) results in an nH value below the Galactic value in the direction of the source (\(5.26 \times 10^{20} \textrm{ cm}^{-2}\); \citealt{2005A&A...440..775K}). So, to correct for this excess soft emission, we added an disk-black body component to our fits (\texttt{Tbabs*(diskbb+powerlaw})). With this model we derive the following parameters: nH = \(0.13 \pm 0.04 \times 10^{22} \textrm{ cm}^{-2}, kT_{bb} = 0.24 \pm 0.03 \textrm{ keV, and } \Gamma= 1.65 \pm 0.08\) with \(\chi2/\nu = 1.00\).

We fitted the \textit{XMM Newton}/EPIC-pn spectrum in the \(0.5-12\) keV energy range in \texttt{XSPEC} using a(\texttt{Tbabs*Powerlaw} model.  With this model we derive the following parameters: nH = \(0.13 \pm 0.04 \times 10^{22} \textrm{ cm}^{-2}, kT_{bb} = 0.24 \pm 0.03 \textrm{ keV, and } \Gamma= 1.65 \pm 0.08\) with \(\chi2/\nu = 1.00\).

%%%%%%%%%%%%%%%%%%%%%%%%%%%%%%%%%%%%%%%%%%%%%%%%%%%%%%%%%%%%%%%%%%%%%%%%%%%%%%%%

\subsection{Hard X-ray}
\label{Hard-X-ray-analysis}

We fitted the \textit{INTEGRAL}/IBIS/ISGRI spectrum in the \(30-400\) keV energy range. A systematic 1.5 \% error was added to the data, following OSA 11 standard recommendations\footnote{https://www.isdc.unige.ch/integral/analysis}.  A power-law fit to the data in \texttt{XSPEC} found a photon index of \(\Gamma=2.41 \pm 0.01\) and a \(\chi^2/\nu = 71.80\).  The spectrum deviates from a simple power-law model, especially at high energies, with residuals suggesting a Comptonized spectrum.  
Fitting the data with a \texttt{CompTT} model using a photon temperature of 0.24 keV fixed to the \(kT_{bb}\) value from \textit{XMM} finds a better fit, with \( k_T = 36.4 \pm 0.9 \textrm{ keV, } \tau = 1.27 \pm 0.05, \textrm{ and } \chi^2/\nu= 6.21\).  When including a reflecting component (\textsc{Reflect}) with the reflection fraction fixed to 1, the fit improves to \( \chi^2/\nu = 3.45 \textrm{ with } k_T = 38 \pm 1 \textrm{ keV and } \tau = 1.44 \pm 0.06\).  Following \cite{2019ApJ...870...92R}, a cutoff power-law was added \texttt{Reflect*(CompTT)+cutoff} with \(\Gamma = 1.6\) and a cutoff energy of 200 keV that improved the fit to \(0.71\) and has fit parameters \( k_T = 27 \pm 4 \textrm{ keV, } \tau = 2.2 \).  

To characterize the X-ray spectrum, a joint fit was performed between the two instruments spanning \(0.5 - 400 \) keV using the model \texttt{Tbabs*Reflect*(diskbb+CompTT)+Tbabs*cutoff} found best-fit parameters of \(kt_{BB} = 0.27 \pm 0.01 \textrm{ keV, } kT = 27 \pm 1 \textrm{ keV, and } \tau = 2.2 \pm 0.1 \textrm{ with } \chi^2 / \nu = 0.95\).

Using this joint spectrum, we calculated the accretion luminosity in the \(1 - 200\) kev energy range and found a value of \( \sim 6 \times 10^{37} \) erg/s for a distance 3 kpc.

\begin{deluxetable}{ccc}
\tablenum{3}
\tablecaption{Irradiated disk fit parameters. \label{table:diskir_para}}
\tablewidth{0pt}

\tablehead{
\colhead{} & 
\colhead{\texttt{diskir}} &
\colhead{\texttt{diskir+po}}
}

\startdata
\( kT_{disk}\)(keV)             & \(0.116 \pm 0.007\)           & \(0.122 \pm 0.007\)             \\
\( \Gamma\)                     & \(1.78 \pm 0.02 \)            & \(1.70 \pm 0.04\)             \\
\( kT_e\)    (keV)              & \(58 \pm 4\)                  & \(37 \pm 4\)                  \\
\( L_C/L_D\)                    & \(4.7 \pm 0.5\)               & \(4.7 \pm 0.6\)                  \\
\( f_{out}\)                    & \((1 \pm 40) \times 10^{-7}\) & \((4 \pm 15) \times10^{-2}\)\\
\( \log(r_{out})\)              & \(3.45 \pm 0.04\)             & \(3 \pm 1\)             \\
\( \Gamma_{po}\)                & \( - \)                       & \(1.6 \pm 0.3\)             \\
Norm\(_{po}\)& \( - \)                       & \(1.0 \pm 1.7\)             \\
\( \chi^2/ \nu\)                &\(1.30\)                       & \(0.97\)                    \\
\enddata
\end{deluxetable}

\subsubsection{IR \(-\) Hard X-ray Spectrum}
\begin{figure}
   \includegraphics[scale=0.6, trim = 20mm 75mm 240mm 152mm]{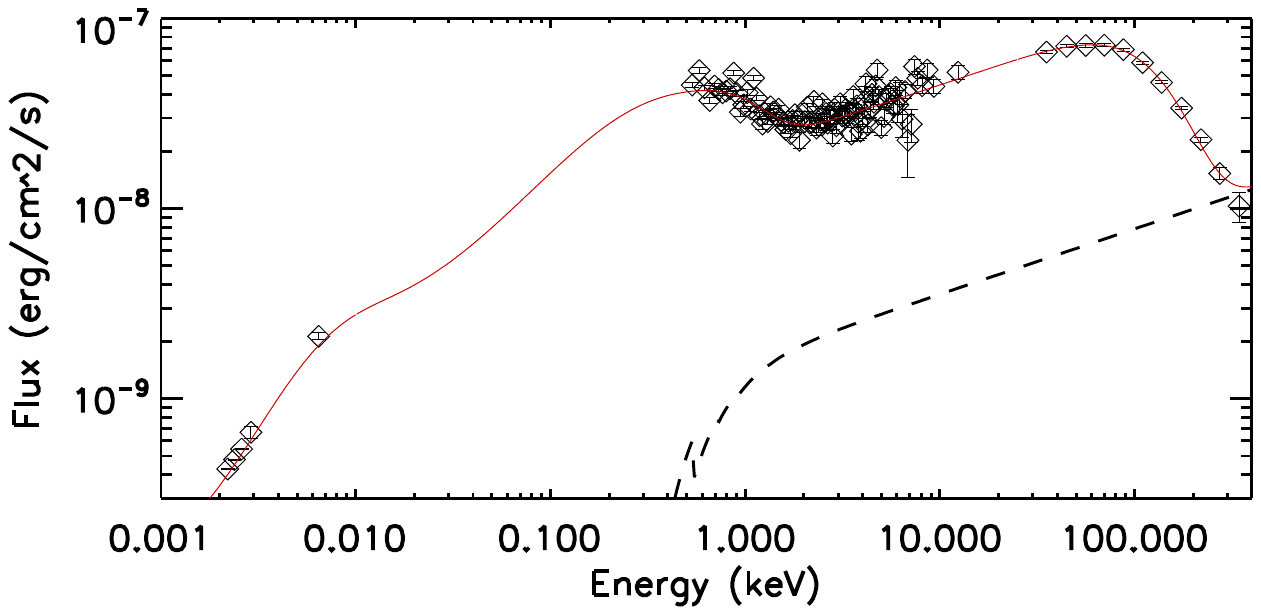}
   \caption{MAXI J\(1820+070\) spectrum from \(0.001 - 400\) keV.  The \texttt{diskir} model is shown with a solid red line and the \texttt{po} component is shown as a black dashed line.}
   \label{fig:diskir_sed}
\end{figure}

Subsequently, we fit our data from the near-IR to hard X-rays using an irradiated disk model to compare with \cite{2018ApJ...868...54S}, which analyzed a similar energy range using observations from 24 March.  The irradiated disk model accounts for the effects of the Comptonized emission on the accretion disk and the soft-excess that is seen in the hard state \citep{DiskIrrModel2009}.  Figure~\ref{fig:diskir_sed} shows the spectrum from \( 0.001 - 400 \) keV with the \texttt{diskir} model shown as a solid red line and the power-law component of the \texttt{diskir+po} model as a dashed black line.  The power law is used to model the high-energy cutoff power-law component in the previous section.  Table~\ref{table:diskir_para} contains the fit parameters using a \texttt{diskir} model with and without an additional power-law component.  We found that including a power-law component improved the fit at high energies and reduced the \( \chi^2 / \nu\) from 1.30 to \(\chi^2 / \nu = 0.97\) with an f-test probability of \(3.7 \times 10^{-8} \).

The origin of the power-law component is unclear.  As shown below in Figure~\ref{fig:jet}, the expected jet flux is too low for the component to be jet emission at those energies.  However, the emission could possibly be from Comptonization of non-thermal electrons as in the case of GRS \(1716-249\) \citep{2020MNRAS.494..571B}.

%In comparison, \cite{2018ApJ...868...54S} found \(kt_{disk} = 0.23 \) keV, \(\Gamma = 1.67 \), \(kT_e = 29 \) keV, \(L_C / L_D = 0.24\), and \(f_{out} = 6.5 \times 10^{-5} \).  Then they included a powerlaw of 1.7 to account for the jet emission in the near-IR.  After which the spectrum was well-described fixing \(kT_{disk} = 0.35 \) keV, \(L_C / L_D = 70 \), \(f_{out} = 5 \times 10^{-3} \), and \(R_{out} = 10^5 R_{in} \). 

%We find a similar power-law index from the near-IR and similar values for \( \Gamma \), \( kT_e \textrm{, and } f_{out}\).  However, \cite{2018ApJ...868...54S} find higher values for \(kT_{disk} \the corona is described by power-law with exponential cutoff'extrm{, } R_{out} \textrm{, and } L_C/L_D \).  

%%%%%%%%%%%%%%%%%%%%%%%%%%%%%%%%%%%%%%%%%%%%%%%%%%%%%%%%%%%%%%%%%%%%%%%%%%%%%%%%

\section{Broadband SED modelling}

\label{Sec:SED}

We modelled the broadband SED of MAXI  J1820+070 using a combination of jet leptonic models and irradiated disk and corona model implemented in Jets SED modeler and fitting Tool (\jetset) \footnote{\url{https://jetset.readthedocs.io/en/latest/}} \citep{jetsetascl,Tramacere2011,Tramacere2009}. A more accurate description of the model is discussed in \cite{TramacereInPrep}.
We assume that the optical/UV up to keV energies is dominated by  disc irradiation and  coronal emission.
The emission in the mm to optical region is dominated by the non-thermal emission of leptons accelerated in the jet by shock and/or stochastic acceleration, and we assume that the break at $\approx 1.5\times 10^{13}$ Hz is due to the transition from the optically thin to the optically thick synchrotron emission.
The radio emission is dominated by the terminal part of the jet that starts beyond the acceleration region and extends up to a distance of $\approx 1\times 10^{15}$ cm according to \cite{2020NatAs.tmp....2B}. A schematic view of the model is provided in Fig.~\ref{fig:jet_diagram}

\subsection{Individual Model Components Description}
In the following we describe the implementation of each model component.

\subsubsection{Irradiated Disk and Hot Corona}   
\label{dikirr-corona-model}
To model the UV to hard-X-ray emission  we have used the disk Comptonization  plus disk irradiation model, \texttt{DiskIrrComp} implemented in \jetset. The \texttt{DiskIrrComp} is based on the
\texttt{diskir} model \citep{DiskIrrModel2009} and the Comptonization model of \cite{Zdziarski2009}.  In detail, we assume that a classical multi-temperature disk with an inner temperature $T_{Disk}$ and an extension $R_{in}=3 R_{S}$ to $R_{out}$, expressed by the dimensionless parameters $r_{in}=R/R_{in}$ and $r_{out}=R/R_{out}$. The disk spectrum is modified due to the reprocessing of irradiated emission from the disk itself and from the corona Compotonization tail.  The corona emission is described by a power law with an exponential cutoff with a photon index $\Gamma_{Comp}$ and a cut-off energy $E_{Comp}=kT_{e}$ where $kT_{e}$ is the corresponding electron temperature. The Compton hump component is described by a power-law with exponential cut-off with a photon index $\Gamma_{hump}$ and a cut-off energy  $E_{hump}$.  We refer to this model as \texttt{Comp. hump}.
The normalization of the Compton tail component is parameterized as a fraction of the disk luminosity $L_{Disk}$ according to $L_{Comp}^{ratio}=L_{C}/L_{Disk}$.  The total bolometric flux will be $L_{bol}=L_{Disk}+L_{rep}+L_{C}$, where $L_{rep}$ represents the thermalized fraction $f_{in}$ of $L_{C}$ thermalized within $r_{in}$ and  $r_{irr}=R_{irr}/R_{in}$, where $R_{irr}$ is the radius of the inner disk irradiated by the Compton tail.
A fraction $f_{out}$ of the bolometric luminosity will irradiate the outer disk. The irradiation creates a shoulder with a spectral trend $f_{out}\propto L_{bol}\nu^{-1}$ that extends between $\nu_1=3kT(r_{out})$ and $\nu_2=3kT(r_{t})$, where $r_{t}$ is the transitional radius between gravitational and irradiation energy release. This effect depends strongly on $r_{out}$ and $f_{out}$, and it is present even without corona Comptonization, because it represents the disk self-irradiation.
The presence of a Comptonization component will provide a further heating of the disk in the inner part modifying the pure gravitational temperature profile.  %To model the presence of a Compton hump we use power-law cut-off model that is provided in \jetset, and we refer to this model as \texttt{Comp. hump}.

\subsubsection{Pre-acceleration and Acceleration Region}    
We assume that electrons in the pre-acceleration region close to the base of the jet are described by a thermal plasma with cooling dominated by adiabatic losses.
Once the particles approach the acceleration region
they are accelerated under the effect of diffusive shock acceleration and/or stochastic acceleration and the corresponding energy distribution   can be modeled by a power-law with a high-energy cutoff
\begin{equation}
  N_{e,acc}(\gamma)=N \gamma^{-s} \exp(-\gamma/\gamma_{cut})
\end{equation}
where the value of $\gamma_{cut}$ takes into account the balance between cooling and acceleration terms. The index $s$ is dictated by the competition of the acceleration time scales and escape time scales \citep{Tramacere2011}. 
We assume that the acceleration region extends from $z_{acc}^{start}$ to $z_{acc}^{end}$,  with cross section   $R_{acc}$ equal to the average cross section of the jet at $z=(z_{acc}^{start} + z_{acc}^{end})/2$, with $z_{acc}^{end} - z_{acc}^{start}=2 R_{acc}$.
The emission from the acceleration region is reproduced using the jet leptonic model \texttt{Jet} implemented in \jetset,  and we refer to it as \texttt{JetAcc} \citep{TramacereInPrep}.

 \subsubsection{Radio Jet}    
 To model the radio jet emission we have used the \jetset  multi-zone radio jet model \texttt{RadioJet}. This model implements a continuous jet as a sum of $N_{c}$ single zones, following the approach of \cite{Kaiser2006}, where for each zone the values of $R$ and $B$ are ruled by Eq. \ref{Eq-jet-ballistic}, and the particle density scales as
\begin{equation}
  N_{s,i}= N_{s,0}(z_{s,0}/z_i)^{m_N}
\end{equation}
where $N_{s,0}$ is the initial density of emitters at the starting point of the radio jet $z_{s,0}$, $z_i$ is the average position of the $i_{th}$ component, and $m_N$ is the index of the particle density law fixed to 2. The initial particle density is a fraction $N_{frac}$ of that present in the acceleration region and we  fix it to 1. The radio jet extends from $z_{radio}^{start}=(z_{acc} +R_{acc}) K_R^{start}$ to  $z_{radio}^{end}(z_{acc} +R_{acc}) K_R^{end}$, where $K_R^{start}$  and $K_R^{end}$ are free parameters. In the present analysis we fix $K_R^{start}= 1$ , and $K_R^{end}$ is fixed in order to match the value of $1\times 10^{15}$ cm according to the analysis presented in \cite{2020NatAs.tmp....2B}.  The particle distribution in each region has the same spectral law as in the acceleration region, but we decrease the value of $\gamma_{cut}$ to take into account the effect of the cooling when the particles leave the acceleration region. In our analysis we take into account only synchrotron cooling and we evolve $\gamma_{cut}$ according to Eq. 27 in \cite{Kaiser2006}. More details about the connection
between the acceleration and radio are discussed in Sec. \ref{model-fit}
%We model the average spectrum of the acceleration region as a single  region, but due to the strong self-absorption, the actual site of  particle injection might extend down to $z_{inj}<z_{acc}$. For this reason we introduce the parameter $z_{inj}^{frac}=z_{inj}/z_{acc}$. 

%and we evolve $\gamma_{cut}$ according
%to \cite{Kaiser2006}:
%\begin{equation}
%\label{radio-synch-cooling}
%\gamma_{cut}=\frac{\gamm_{inj}}{1+\fra{4\sigma_T B^2}{3 m_e c 8 %\pi}}\gamma_{inj}t_0^{2 m_B}
%    
%\end{equation}

%We ignore the evolution of $\gamma_{cut}$ along the radio jet because the %decreasing of the magnetic field, and the decreasing of the adiabatic %cooling rate will not change significantly the position of $\gamma_{cut}$, %moreover, the largest fraction of the emissivity is due to the turn-over %region of the synchrotron spectrum, where the radiation is emitted by  %electrons with a Lorentz factor of a few. A self-consistent implementation %and description of this model that is beyond the scope of this analysis %will be presented \cite{TramacereInPrep}.

\begin{figure*}
    \centering
    \includegraphics[width=0.8\linewidth]{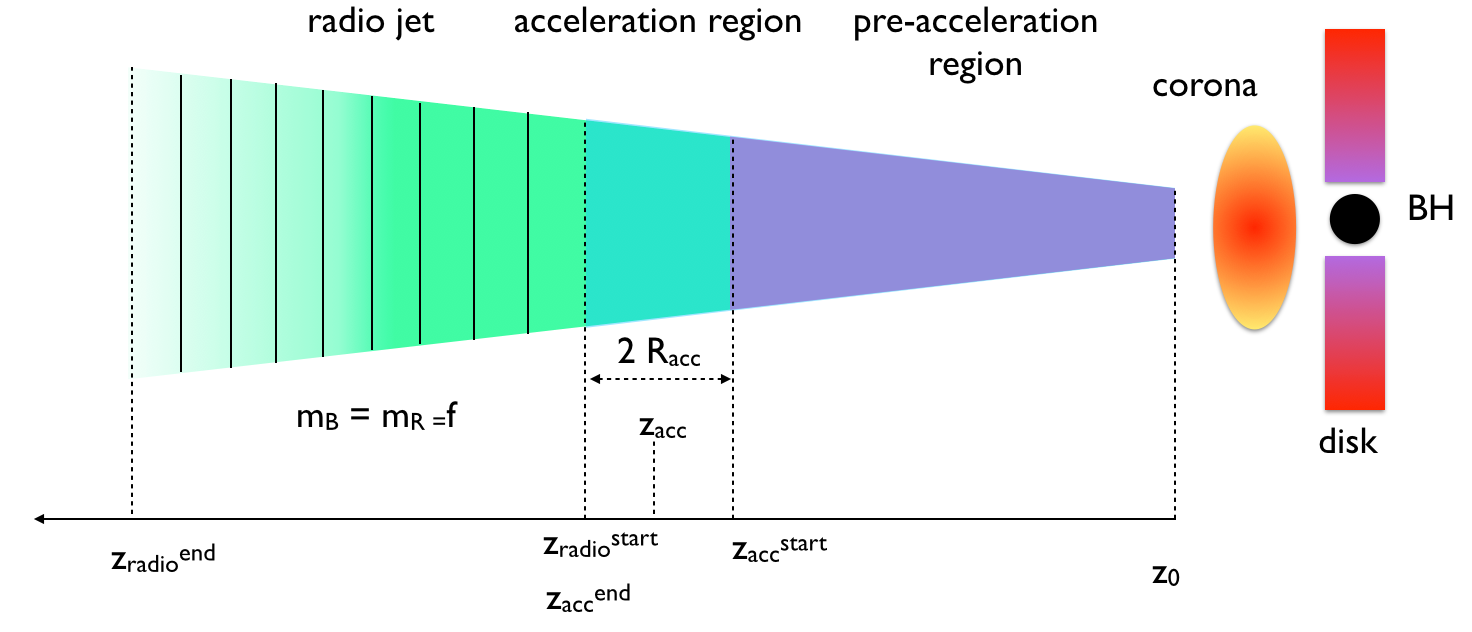}
%    \caption{}
    \caption{A schematic view of the jet model setup. The purple region identifies the pre-acceleration region, the cyan region identifies the acceleration region, and the green region identifies the radio jet. The $z$ axis on the bottom shows the starting and end end point of each region. The acceleration region is assumed to be spherical with a radius equal to the jet cross section. The vertical black lines in the radio jet region marks qualitatively the division of region in slices.}
    \label{fig:jet_diagram}
\end{figure*}

\subsection{Phenomenological Model Setup}
\label{model-setup}
As a first step we set the geometrical properties of the jet, i.e. we define the extent of the pre-acceleration, acceleration, and radio emission sites, and the values of the magnetic field.
We assume that the jet is launched at distance $z_0$ from the BH, with an initial cross section $R_0$, and that the bulk Lorentz factor ($\Gamma_{jet}$) of the jet is constant over the full jet extent.  The acceleration region starts at a distance $z_{acc}^{start}$ with a width equal to jet cross section  diameter $R_{acc}=2R(z_{acc})$, and we treat it as a spherical region. The radio region starts at $z^{start}_{radio}=K_{R}^{start}( z_{acc} + R_{acc}$) and ends at a distance $z^{end}_{radio}=K_{R}^{end} (z_{acc} +R_{acc})$ (a scheme of the model is presented in Fig. \ref{fig:jet_diagram}).
According to \cite{2020NatAs.tmp....2B} we fix the distance of jet from the observer to the value of $d=3$ kpc, the termination of the radio jet to the value of  $z_{end}\approx 1\times 10^{15}$ cm, and the value of the beaming factor to $\delta=[\Gamma_{jet}(1-\beta_{jet}\cos(\theta_{obs})]^{-1}\approx 2.2$, using the values of  $\beta_{jet}=0.89$ and $\theta_{obs}= 63^{\circ} $ reported in \cite{2020NatAs.tmp....2B}. We assume a ballistic jet model \citep{Kaiser2006,BBR} characterized by
\begin{eqnarray}
    B(z) &\propto& B_0(z_0/z)^{m_B} \nonumber \\ 
   \label{Eq-jet-ballistic}
    R(z) &\propto& R_0(z/z_0)^{m_R} \\ 
%    \notag 
    N(z) &\propto& R_0(z_0/z)^{m_N}  \nonumber 
\end{eqnarray}

with $m_B\approx 1$ and $m_R = 1$, and $m_N=2.0$. This choice assumes that the jet is very close to the ballistic regime, with a magnetic field dominated by the toroidal component and justifies the assumption that the bulk Lorentz factor is constant along the jet.

We use a black hole mass of $M_{BH}= 8 M_{\Sun}$.  The jet luminosity $L_{jet}$ is linked to the Eddington luminosity ($L_{Edd}$) according to
\begin{equation}
 L_{jet}=  \frac{1}{2} q_{jet} L_{Edd}
 \label{Eq-jet-eff}
    \end{equation}
where $L_{Edd}\approx 1.3\times 10^{38} (M_{BH}/M_{\Sun})$ erg s$^{-1}$ \citep{Ryb1986}.
It is worth noting that our $q_{jet}$ parameter is not  linked directly to the accretion efficiency process, because the jet powering could, in principle, be supported also by other mechanisms such as the Blandford–Znajek mechanism \citep{BZ1977}, that predicts electromagnetic extraction of energy and angular momentum  from magnetized accretion disc surround a black hole. Hence, our $q_{jet}$ parameter should not be used to infer or constrain the accretion efficiency, and will be discussed in a more accurate physical context in \cite{TramacereInPrep}.

We assume that the jet is launched at a  distance $z=z_{0}$ from the BH  with $z_0=50 R_S\approx 1.2 \times 10^{8}$ cm, where $R_S=(2GM_{BH})/{c^2}$.
The launching jet position $z=z_{0}$, in the current analysis is assumed constant, to reduce the model complexity, and it is chosen according to reference values published in previous analysis \citep{Romero2010}. 
The initial radius of the jet is set to $R(z_0)=0.1 z_0$, resulting in an opening angle of $\theta_{open} \approx 5.7 ^\circ$.
We impose that in the launching region the entire jet power is in the form of magnetic energy
\begin{equation}
     L_{jet}= L_B(z_0)= \pi U_B(z_0) R(z_0)^2\Gamma_{jet}^2\beta_{jet}c
    \label{Eq-jet-eff}
\end{equation}
where $U_B=B^2/(8\pi)$, and setting $q_{jet}=0.2$ we obtain $B_0\approx 6.8 \times 10^{6}$ G.

%\begin{figure}
%    \centering
%    \includegraphics[width=8.5cm]{figures/radio_opt_Fnu_plot.pdf}
%    \caption{Jet emission between radio and optical bands used for the %phenomenological model setup. The black dashed line is a phenomenological %approximation (Eq. 1 in  \cite{Stevens1995}) \note{of the optically-thin to %optically-thick synchrotron  emission from  a power-law cut-off electron %distribution in the putative acceleration region, plus a power-law spectral %emission accounting for the Optical/UV rising part of the spectrum.  The %black dashed line is a power-law fit of the radio jet region.}}
%    \label{fig:radio_opt_Fnu_plot.pdf}
%\end{figure} 

The value of $m_B$ can be constrained from the spectral index of radio jet emission, $\alpha_{R}\approx 0.15$, according to the Eq. 39 in \cite{Kaiser2006} that refers to the case of strong radiative cooling 
and almost constant value of the electron distribution high-energy cutoff. According to this scenario, that is very similar to what we expect in our case, we can rearrange Eq. 39 in \cite{Kaiser2006} as: 
\begin{equation}
    m_B=\frac{1+m_R}{2 - \alpha_{R}}
\end{equation}
that is similar to the trend of the thick radio spectrum discussed in 
\cite{Peer:2009},  and we obtain a value of $m_B\approx 1.1$
We stress that this is an initial guess done assuming that 
the jet is not changing after the acceleration region. As we will discuss in the next section, during the model fit we need to take into account that jet expansion might change above the acceleration region, hence we will relax the constraint on $m_{B}$ and $m_R$ considered in the \texttt{RadioJet} emission.

To constrain the value of $z_{acc}$   we impose that 
$R_{acc} =R(z_{acc} )$, $B_{acc} =B(z_{acc} )$  and $N_{e,acc}$  correspond to a synchrotron self-absorption frequency of  $\nu_t \approx 1.5 \times10^{13}$ Hz. This value of $\nu_t$ is obtained from the phenomenological fit of  the optically-thin to optically-thick synchrotron emission between mm and optical data shown in  Fig. \ref{fig:jet_compact}. In order to solve this problem we combine the analytical expression of the synchrotron self-absorption frequency ($\nu_t$) \citep{Ryb1986}, evaluated at the peak i.e. $\alpha_{\nu}=0$
\begin{equation}
    \nu_t=\nu_L 
    \Big[
    \frac{\pi\sqrt{\pi}}{4}
    \frac{q R_{acc}  N_{e,acc}}{B_{acc} }
    f_{k}(s)
    \Big]^{\frac{2}{s+4}},
     \label{eq-synch-nu-abs}
\end{equation}
and of that the synchrotron emissivity  \citep{Ryb1986} $\epsilon_s(\nu)$ :
\begin{equation}
    \epsilon_s(\nu)= \frac{F_\nu d_L^2 }{V}= 
    \frac{3\sigma_T c N_{e,acc} U_B^{acc}}{16 \pi \sqrt(\pi) \nu_L}
    f_{\epsilon}(s),
    \label{eq-synch-em}
\end{equation}
where $q$ is the electron charge, $ U_B^{acc}$ is the value of the magnetic field, $V$ is the volume of a spherical geometry of volume $V$ of radius $R_{acc}$, $s$ is the slope of the electron distribution power-law and $\nu_L=\frac{qB}{2\pi m_e c}$ is the
Larmor frequency, and where the functions $f_{k}(s)$  and $f_{\epsilon}(s)$ are approximated to  percent accuracy   as reported in \cite{Ghisellini2013}. 
The value of $s$ is obtained using the optically thin spectral index $ \approx 0.6$ from the phenomenological fit in Fig. \ref{fig:jet_compact}, according
to the relation $s= 2\alpha +1\approx 2.2$ \citep{Ryb1986}.
We solve  Eq. \ref{eq-synch-em} with respect to $N_{e}^{acc}$ and then substitute in Eq. \ref{eq-synch-nu-abs}, and we insert the functional form of $B=B(z_{acc})$ and $R=R(z_{acc})$ according to Eq. \ref{Eq-jet-ballistic}. The final equation solved with respect to $z_{acc}$ reads:

\begin{equation}
z_{acc} = \Big[
\Big(\frac{\nu_t 2\pi m_e c}{q B_0^2 z_0^{m_B}}\Big)^{\frac{s+4}{2}}
\frac{3 \sigma_{T}   (B_0 R_0)^2  f_{\epsilon}(s)  z_0^{2 \Delta_m}}{16 r_e^2 \pi^3   f_k(s) F_{\nu} d_L^2  }
\Big]^{\psi}
\label{eq-z_acc}
\end{equation}

where $r_e=q^2/(m_e c^2)$ is the classical electron radius,  $\Delta_m  =   m_B - m_R $, and $\psi   =  \frac{2}{ 4  \Delta_m  - m_B (s + 4 )}$.
%\begin{eqnarray}
% \psi  & = & \frac{2}{ 4  \Delta_m  - m_B (s + 4 )} \\
% \Delta_m  & = &  m_B - m_R ,  \nonumber
%\end{eqnarray}

%\note{(AT) We set the value of the power-law slope electron distribution \note{to $s\approx 2.2$},  and that of  $F_{\nu}^t\approx 0.5$ Jy.  Both these values are derived from the phenomenological fit in Fig. \ref{fig:jet_compact}. The luminosity distance is set to the  estimated value of $d_L = 3$ kpc. 
%In this way we obtain   \note{$z_{acc}\approx 2.8 \times 10^{10}$ cm corresponding to $\approx 1.4 \times 10^2 R_{S}$, with $R_{acc}\approx 2.8\times 10^{19}$ cm,   a value of  $B_{acc}\approx 1.6 \times 10^4$  G,
%and $z_{acc}^{start}\approx  \times 2.5^{10}$ cm}

Consequently, the starting position of the radio jet is set to  $z_{radio}^{start}=z_{acc}^{end}=z_{acc}+R_{acc} \approx 3.1\times 10^{10}$ cm, with an extent derived from \cite{2020NatAs.tmp....2B} of $z_{end} \gtrsim 30000 z_{radio}^{start}$

The value of the cut-off of the electron distribution is set to $\gamma_{cut}=60$, in order to produce the peak of the synchrotron emission above the IR frequencies for a magnetic field $B_{acc}\approx 1.8 \times 10^4$ G, with a power-law slope $s\approx 2.1$ that is slightly lower then the value derived from the optically thin spectral index.

The constrained value of $z_{acc}$ can be used to derive the hadroninc content of the jet energetic in form of cold protons. 
Following \cite{Romero2010} we impose that in the acceleration 
region of the jet the magnetic energy of the jet is in subequipartition with the bulk kinetic energy of the cold protons, a condition that is mandatory to allow the mechanical compressibility of the plasma \citep{Kossimarov2007}. 
We define the parameter $\rho^{acc}_{p,B}=U_{p}(z_{acc})/U_{B}(z_{acc})$, where $U_{p}(z)=n_p(z) m_p c^2$, and we require that $\rho_{p,B}>1$. This choice sets a value of cold proton luminosity in the acceleration region $L_p(z_{acc})>3.6 \times 10^{37}$ erg $s^{-1}$.  
%We set a value of $\rho^{acc}_{p,B}=5.0$ resulting in $L_p(z_{acc}) \approx 2.5\times 10^{38}$ erg $s^{-1}$.

%A resume of the starting values of the model is reported in Tab. \ref{best-fit-table} in the column `staring values'.

\begin{center}
\begin{table}
\centering
\tablenum{4}
\label{phenomenological-table}
\caption{Phenomenological Setup  Parameters} 
\begin{tabular}{lcc}
\hline
\hline
%\multicolumn{3}{c}{phenomenological setup  parameters}                                                  \\
\multicolumn{3}{c}{                                  }                                                  \\

\hline
\multicolumn{3}{c}{ Input Parameters}                                                                   \\
\hline    
  par. name             & units                            &  input value                                \\
\hline
 $z_0$                    &  cm  				           &  $1.12   \times 10^{8}$				     \\
 $r_0$                    &  cm  				           &  $1.12   \times 10^{7}$				     \\
 $M_{BH}$                 &  $M_{\Sun}$   			       &  8 				                         \\
 $q_{jet}$                &    				               &  0.20				                         \\
 $F_{\nu}^t$              &  Jy                            &  $0.5 $						             \\
 $\nu_t$                  &  Hz                            &  $1.5   \times 10^{13}$ 				     \\
 $s$                      &                                &  2.1					                     \\
 $\rho^{acc}_{p,B}$       &                                &  $>1$                                       \\
 $m_{B}$                  &                                &  1.1	                                     \\
 $m_{R}$                  &                                &  1.0					                     \\

\hline
\multicolumn{3}{c}{Output Parameters}                                                                    \\
 \hline    
  par. name               & units                          &  output value                               \\
\hline
 $B_0$                    &  G                             &  $6.8     \times 10^{6}$                    \\
 $B_{acc}$                &  G                             &  $1.8\times 10^{-4}$                                     \\
 $L_{p}^{acc}$            &  erg s$^{-1}$                  &  $>3.6     \times 10^{37}$       		     \\
 $z_{acc}^{start}$        &  cm  				           &  $2.4      \times 10^{10}$					 \\
 $z_{acc}^{end}$          &  cm  				           &  $2.9      \times 10^{10}$					 \\
 $z_{acc}$                &  cm  				           &  $2.6      \times 10^{10}$					 \\
 $R_{acc}$                &  cm  				           &  $2.6      \times 10^{9}$					 \\
 $z_{radio}^{start}$      &  cm  				           &  $2.9      \times 10^{10}$					 \\
 $z_{radio}^{end}$        &  cm  						   &  $\gtrsim 1\times 10^{15}$	                 \\
\end{tabular}
\end{table}
\end{center}

\begin{center}

\begin{table*}
\centering
\tablenum{5}
\label{best-fit-table}
\caption{\jetset best fit model parameters }

\begin{tabular}{clcccccc} 
\hline
\hline
\multicolumn{8}{c}{}  
%\multicolumn{8}{c}{ }                                                                                                                                                 \\ 
                                                                                                                                               \\ 
\hline
model name           & par. name           & units         & best fit value            &error               & starting value        & fit boundaries                                & frozen   \\ 
\hline          
\texttt{CompHump}    & $E_{hump}$          & keV           & 26                        & 14                   & 20                    & {[} 15 ; 35]                                  & False   \\
"                    & $\Gamma_{hump}$     &               & -0.5                      & 2                    & -1.2                  & {[} -2 ; 2]                                   & False   \\
%"                   & $K_{hump}$          &               & $9.9\times 10^{-4}$       &$7\times 10^{-6}$     & $1.5\times 10^{-4}1$  & {[} $5\times 10^{-5}$ ; $2\times 10^{-4}$]    & False   \\
%"                   & $\alpha_{hump}$     &               &                           &                      & 1                     &                                               & True    \\
\texttt{DiskIrrComp} & $T_{Disk}$          & K             &                           &                      & $1.55 \times 10^6$    &                                               & True    \\
"                    & $ L_{Disk}$         & erg s$^{-1}$  & $1.09\times 10^{37}$       & $1.0\times 10^{32}$ & $1 \times 10^{37}$    & {[} $1\times 10^{36}$ ; $ 1 \times 10^{39}$]  & False   \\
%"                   & $\theta$            & deg           & 54                        & 1                    & 60                    & {[} 50 ; 70 ]                                 & False   \\
"                    & $r_{out}$           &               & $3.58\times 10^{3}$       & $0.21\times 10^{3}$  & $5 \times 10^{3}$     & {[} 1 ; – ]            & False   \\
"                    & $r_{irr}$           &               &                           &                      & 1.1                   &                                               & True    \\
"                    & $\Gamma_{Comp}$     &               & 1.64                      &0.12                  & 1.65                  & {[} 1.3 ; 1.9 ]                               & False   \\
%"                   & $\alpha_{Comp}$     &               & 0.90                      &0.01                  & 1.0                   & {[} - ; - ]                                   & False   \\
"                    & $E_{Comp}$          & keV           & 150                       &100                   & 140                   & {[} 20 ; 200 ]                                & False   \\
"                    & $L_{Comp}^{ratio}$  &               & 4.1                       &0.6                   & 4.5                   & {[} 0 ; – ]                                   & False   \\
"                    & $f_{in}$            &               &                           &                      & 0.1                   &                                               & True    \\
"                    & $f_{out}$           &               & $1 \times 10^{-2}$        &$40 \times 10^{-2}$   & 0.01                  & {[} 0 ; – ]                                   & False   \\
\hline 	                          
\texttt{DiskIrrComp} & $r_{out}$           &               & $3.4\times 10^{3}$        &$0.5\times 10^{3} $   & $3.58\times 10^{3}$   & {[} 1 ; – ]           & False   \\
"                    & $f_{out}$           &               & $7.33\times 10^{-3}$      &$0.15\times 10^{-3}$  & $1 \times 10^{-2}$    &  {[} 0 ; – ]                                    & False   \\
"                    & $L_{Comp}^{ratio}$  &               & 4.270                     &0.016                 & 4.1                   &  {[} 0 ; – ]                                   & False   \\
\texttt{JetAcc}      & $N_{e,acc}$         & cm$^{-3}$     & $9.998\times10^{11}$      & $0.001\times10^{11}$ & $1.0\times10^{12}$    &  {[}0 ; – ]                                    & False   \\
"                    & $s$                 &               & 2.082                     & 0.007                & 2.1                   &  {[}- ; - ]                                    & False   \\
"                    & $\gamma_{cut}$      &               & $65.4$                    & $1.7$                & 60                    &  {[}1 ; – ]                                    & False   \\
"                    & $R_{acc}$           & cm            & $2.6\times 10^{9}$        & $1.0\times10^{1}$    & $2.6\times 10^{9}$    &  {[} $1.32\times 10^{9}$ ; $ 3.96 \times 10^{9}$]   & False   \\
"                    & $z_{acc}$           & cm            &                           &                      & $2.8\times 10^{10}$   &                                               & True    \\
"                    & $B_{acc}$           & G             & 17986                     & $1.0\times10^{-3}$   & 17986                 & {[} 8993 ; 26980 ]                            & False   \\
"                    & $\theta_{jet}$      & deg           &                           &                      & 63                    &                                               & True    \\
"                    & $\Gamma_{jet}$      &               &                           &                      & 2.19                  &                                               & True    \\
\texttt{RadioJet}    & $z_{inj} $          &               &                           &                      & $2.5\times 10^{10}$   &                                               & True   \\
"                    & $N_{frac}$          &               &                           &                      & 1                     &                                               & True    \\
"                    & $K_{R}^{start}$     &               &                           &                      & 1                     &                                               & True    \\
"                    & $K_{R}^{end}$       &               &                           &                      & 30000                 &                                               & True    \\
"                    & $m_{jet}$           &               &  1.203                    &0.001                 & 1.1                   &  {[} 0.5 ; 1.5 ]                              & False   \\
%"                   & $m_N$               &               &                           &                      & 2                     &                                               & True    \\
\hline
\end{tabular}

\end{table*}
\end{center}

%\begin{figure*} 
%    \centering
%     \includegraphics[width=1.0\linewidth]{figures/jetset_best_%fit_model_1_nuFnu}
%    \caption{Same as in Fig. \ref{fig:jet} for the $\nu %F_{\nu}$ representation of the global model fit }
%    \label{fig:jet_nuFnu}
%\end{figure*}

\subsection{Model Fit and Results}   
\label{model-fit}
\subsubsection{Initial model setup}
To optimize the model we use the composite model interface \texttt{FitModel} provided by \jetset, that allows combining different models in a global model. This model can be optimized by inserting it to the \texttt{ModelMinimizer} \jetset plugin. In the current analysis we use a frequentist approach and we use the \texttt{Minuit} ~\texttt{ModelMinimizer} option.
We have used the ~\texttt{Data} and ~\texttt{ObsData} \jetset tools to import the observed data, and we have added a 5\% systematic error in the range $[1\times 10^{8},1\times 10^{16}]$ Hz,  to avoid that the large inhomogeneity on the fractional error between radio and  optical/UV data, could bias the fit convergence. For the error estimate we provide only errors derived form the \texttt{MIGRAD} module of \texttt{Minuit}, a more reliable estimate based on a Markov chain Monte Carlo (MCMC) will be presented in \cite{TramacereInPrep}

The \texttt{DiskIrrComp} model, the \texttt{Comp. hump} model, and the \texttt{JetAcc} are independent, on the contrary, \texttt{JetAcc} and
radio \texttt{RadioJet} are bound.

\begin{figure*} 
\centering
    %\begin{subfigure}
    \includegraphics[width=0.83\textwidth]{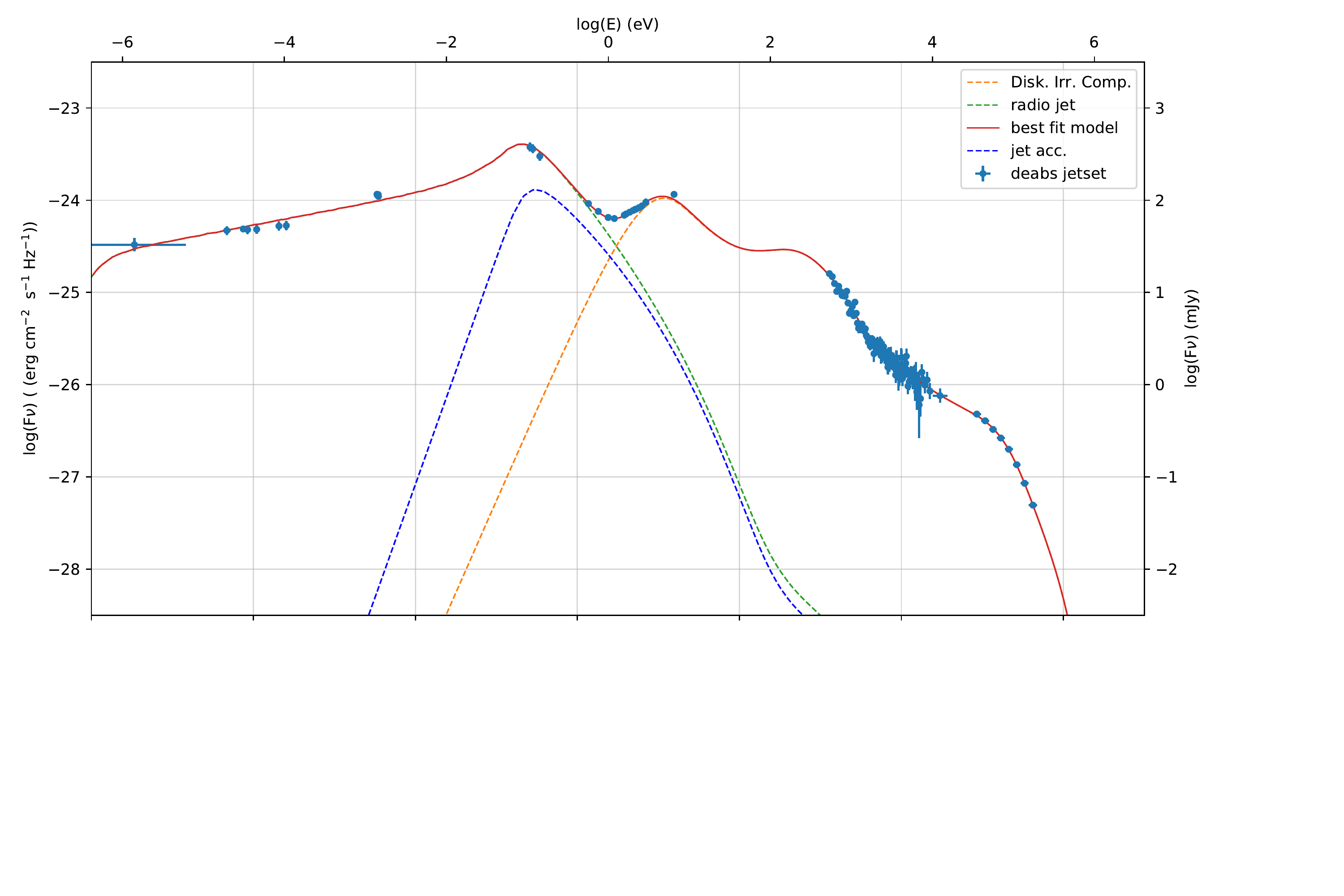}\\
    \vspace*{-23mm}

%\end{subfigure}
        %\begin{subfigure}
     \includegraphics[width=0.83\textwidth]{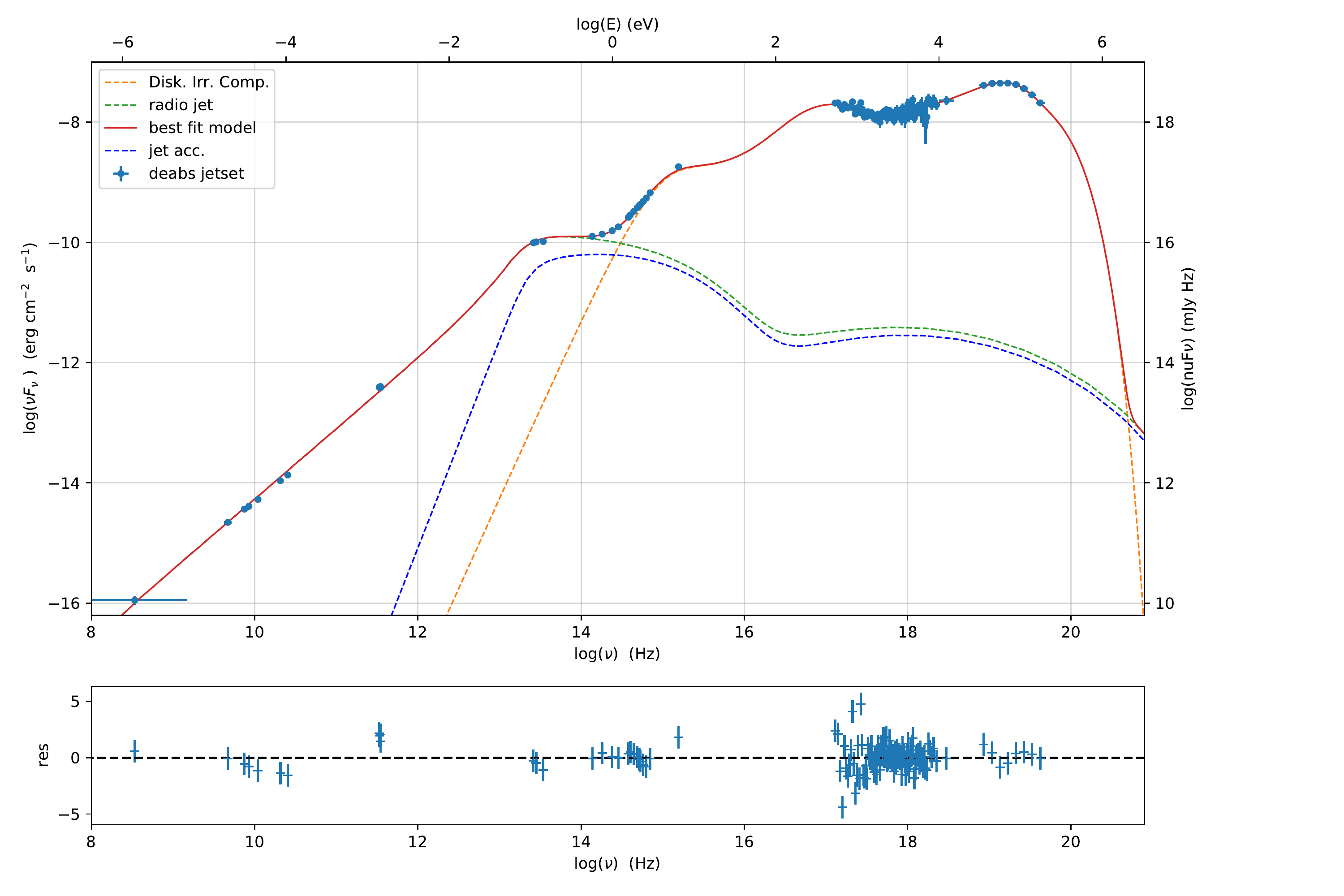}
    %\end{subfigure}

    \caption{The best-fit \jetset model of the broadband SED.    Top panel: the  $F_{\nu}$ representation of the global model fit. Bottom panel: the $\nu F_{\nu}$ representation.
    The red line 
    represents the global model, the dashed lines correspond to the single components, the color is reported in the legend. The best fit parameters
    are reported in Tab. \ref{best-fit-table}. The residuals plot is evaluated with respect to the $\nu F_{\nu}$ representation. }
    \label{fig:jet}
\end{figure*}

The  initial values of the parameters for the \texttt{DiskIrrComp} model are chosen according to the analysis presented in Sec~\ref{Soft-X-ray-analysis} and Sec.~\ref{Hard-X-ray-analysis}. In detail, we set the initial values of $L_{disk}=1\times 10^{37}$ erg s$^{-1}$, of  $r_{out}=5000$, of $f_{out}=0.01$, and $L_{Comp}^{ratio}=4.5$ and we fix the inner disk temperature to $T_{Disk}=1.55\times 10^{6}$K, and the parameters $r_{irr}=1.1$ and $f_{in}=0.1$, the choice adopted in \cite{DiskIrrModel2009}, when the Comptonization of the outer disk is included in the irradiated disk.

For the \texttt{JetAcc} model, we fix $\theta_{jet}=63^{\circ}$, $\Gamma_{jet}=2.19$, we  put a relative bound of +/- 0.5 centered on the parameters values derived in the previous section, $R_{acc}=2.6 \times 10^{19}$ cm, and $B_{acc}\approx 1.8 \times 10^4$ G,  we freeze the initial value of $z_{acc}=2.6 \times 10^{10}$ cm, 
and we leave free the parameters for the electron distribution.

%\subsubsection{connetiont of the acceleration regio to the radio jet}
The initial setup of the parameters of the  \texttt{RadioJet} is more complex and we need to take into account the physical connection with the acceleration region and the cooling process.
%The most important effect is due to the evolution of $\gamma_{cut}$. In this regard we note that, even though the spectrum of the acceleration region is modeled as a single region, due to the strong self-absorption, the actual site of  particle injection might extend down to $z_{inj}<z_{acc}^{start}$, where we parametrize $z_{inj}$ as $z_{inj}=z_{inj}^{frac} z_{acc}^{star}$.  
This effect plays a crucial role, indeed, as already discussed in \cite{Kaiser2006} and \cite{Peer:2009}, the combination of synchrotron cooling and jet expansion (assuming a negligible contribution from adiabatic cooling) will result in an asymptotic value of $\gamma_{cut}(t)$, that can naturally explain the flat radio spectrum without the need to introduce significant particle re-acceleration in the radio jet. We follow the approach reported in \cite{Peer:2009} (in the case of negligible adiabatic cooling) and we set $m_B^{radio}=m_R^{radio}=m_{jet}$.
The particle cut-off evolution in the radio jet will evolve according to \citep{Kaiser2006}:
\begin{equation}
\label{radio-synch-cooling}
\gamma_{cut}(t)=
\frac{\gamma_{cut}}
     {1+
     \frac{\sigma_T B0^2}
          {6 m_e c\pi (f) }
    \gamma_{cut}
    t_0^{1-f}
    (t^{f} -t_{inj}^{f})}
\end{equation}

where $f=1 - 2 m_{jet}$, and $t_0=z_0/\beta_{jet}c\Gamma_{jet}$, $t_{inj}=z_{inj}/\beta_{jet}c\Gamma_{jet}$   and $t=z_/\beta_{jet}c\Gamma_{jet}$, are the comoving time scales. We freeze a starting value of $z_{inj}=z_{acc}^{start}\approx 2.5 \times 10^{10}$ cm.

Another effect to take into account is the fact that for $z>z_{acc}$ the structure of the jet could change, for this reason we leave free the parameters $m_{jet}$ with a fit boundary of [0.5,1.5], with an initial value of 1.18, that is slightly larger than the value used for the phenomenological constraining, but gives a better agreement with radio-to-optical data.

The density of emitters at the base of the \texttt{RadioJet}, is bound to be equal to the density of emitters in the acceleration region $N_e$ calculated according to Eq. \ref{Eq-jet-ballistic}, at $z=z_{radio}^{start}$, by fixing $N_{frac}=1.0$. We fix the  values of $K_{R}^{end}=3000$ and  of $K_{R}^{start}=1$.
    
A list of the free and frozen and of the bounds  is reported in Table~\ref{best-fit-table} in the columns `staring values',  `fit boundaries', respectively.

\subsubsection{Model  fit results for the Disk and Corona emission}

We fit first the \texttt{DiskIrrComp} and  \texttt{Comp. hump} components restricting the fit range to $\nu$= $[5\times 10^{14},10^{20}]$ Hz and we get  $\chi^2 =152 $ for 98 degree of freedom ($N_{dof}$), corresponding to reduced $\chi^2_{red}=1.55$. The parameters values are reported in the upper part of Table~\ref{best-fit-table}.

The best-fit parameters resulting from \jetset are similar to those obtained form the  \texttt{XSPEC} analysis for the \texttt{diskir} model. In particular the  \(r_{out}\) value  (\(3.58 \times 10^3 R_{in} \) and \(3.45 \times 10^3 R_{in}\)) and \(L_C / L_D \) (4.1 and 4.6), for \jetset and \texttt{XSPEC} respectively. The \(f_{out}\) parameter, is unconstrained both for \texttt{XSPEC} and the \texttt{JetSeT}. However, a well constrained value is obtained when the jet component is added as shown in section \ref{model-fit-jet}.
Because the \texttt{JetSeT} model for the Comptonized emission is phenomenological, the high-energy range of the irradiated disk is fit as a cutoff power-law and thus is not directly comparable to the \texttt{diskir} parameters.  For that portion of the spectrum, \texttt{JetSeT} found \( \Gamma = 1.64\) (compared to 1.78 from \texttt{diskir}) and \(E_C = 150 \) keV (compared to \(kT_e = 58\) keV from \texttt{diskir}).  We do note that \(E_C / kT_e \approx 2.6\), which falls within the predicted range of ratios between cutoff energy and electron temperature \citep{2000ApJ...540..131P,2001ApJ...556..716P}, suggesting that the values are in agreement, even though the incertitude on the \texttt{JetSeT} value is quite large.

\subsubsection{Model fit results for the jet emission}
\label{model-fit-jet}
To fit the full band SED  we freeze all the parameters in the \texttt{CompHump} and \texttt{DiskIrrComp} components, except for   \(r_{out}\),  \(f_{out}\), and \(L_C / L_D \), and we fit the global model  over the full SED band in the range $\nu$= $[5\times 10^{8},10^{20}]$ Hz 
.

The  model fit converged with a final $\chi^2 = 181 $ for 122 degree of freedom ($N_{dof}$), corresponding to a $\chi^2_{red}=1.48$. The parameters values are reported in the bottom part of Table \ref{best-fit-table}, and the 
parameters derived from the best-fit model are reported in Table \ref{model-fit-derived-table}.
Regarding the \texttt{DiskIrrComp}, we note that adding the jet component results in a better constraint on the value of  \(f_{out}\)=$(7.33\pm0.15)\times10^{-3}$, that  is in the expected range of other black hole binaries in the hard state \citep{DiskIrrModel2009}.
Moreover, restricting the fit statistics to the same interval used in the  \texttt{XSPEC}  analysis, we get a  $\chi^2 = 157 $ with $N_{dof}=107$,   corresponding to a $\chi^2_{red}=1.6$. 
Regarding the jet component, we note that final best-fit model parameters 
did not change significantly from the input values, suggesting that the phenomenological setup was able to find a configuration very close to the optimal one, even though the fit might be biased by the degeneracy among some parameters. We will investigate this problem in a forthcoming work \cite{TramacereInPrep}, by means of Bayesian approach based on the MCMC technique.
In general, we find that our assumption based on the connection between a compact acceleration region feeding the extended radio jet is able to model self-consistently the UV-to-optical emission, reproducing the observed flat radio spectrum. In  particular, we find that, according to our best fit model, the particles in the radio region reach the asymptotic value of $\gamma_{cut} \approx 8$, and keep it almost constant as result of the decrease in the cooling synchrotron cooling rate due to the jet expansion. This behaviour is in agreement with the results of \cite{Peer2009} and \cite{Kaiser2006} for the case of synchrotron cooling dominating over the adiabatic one.
It is worth discussing some specific parameters in detail:

\begin{table}
\centering
\tablenum{6}
\label{model-fit-derived-table}
\caption{Model parameters evaluated from the best-fit model}
\begin{tabular}{lccc} 
\hline
\hline
%\multicolumn{4}{c}{}                                     \\
\multicolumn{4}{c}{}                                     \\

\hline    
  par. name             & units                 & value                          &  setup value             \\
\hline
 $q_{jet}$              &    				    & $>0.15$                        &  0.20				    \\
 $U_e/U_B$              &    				    & 0.18                            &  --						\\
 $N^{acc}_e/N_p^{cold}$ &    				    & $<94$                         &  --						\\
 $L_{jet}^{acc}$        &  erg s$^{-1}$         & $>8.0  \times 10^{37}$         &  --						\\
 $L_{rad}^{acc}$        &  erg s$^{-1}$         & $1.1   \times 10^{36}$         &  --						\\
 $L_{B}^{acc}$          &  erg s$^{-1}$         & $3.6   \times 10^{37}$         & $3.6 \times 10^{37}$	    \\
 $L_{e}^{acc}$          &  erg s$^{-1}$         & $6.6   \times 10^{36}$         &  --					    \\
 $L_{p}^{acc}$          &  erg s$^{-1}$         & $>3.6  \times 10^{37}$         & $>3.6 \times 10^{37}$    \\
\hline
\end{tabular}
\end{table}

\begin{itemize}
    \item $q_{jet} > 0.15$. This value is compatible with the input value $q_{jet}=0.2$. As already discussed in the previous section, the $q_{jet}$ parameter is not  linked directly to the accretion efficiency because the jet powering could, in principle, be supported also by other mechanisms such as the Blandford–Znajek mechanism \cite{BZ1977}, that takes into account advection of magnetic flux from an accretion disk surrounding the Black Hole. Hence, our $q_{jet}$ parameter can not be used to infer or constrain the accretion efficiency.

    \item $U_e/U_B=0.18$. The  $U_e/U_B$ is not far from equipartition, and it is obtained  without providing any constraint. This proves that the combination of the phenomenological model setup and the minimization of the global model converged naturally toward a configuration close to the physical equipartition of $U_e$ and $U_B$, giving further support to the choice of a compact acceleration region that is connecting the pre-acceleration region to the radio jet.
    
    \item $N^{acc}_e/N_p^{cold}<94$.  
    Since our model is leptonic, the content of cold protons can be derived from ancillary conditions, as the condition  that  the magnetic energy of the jet has to be in subequipartition with the bulk kinetic energy of the cold protons, in order  to allow the mechanical compressibility of the plasma \citep{Kossimarov2007}, and formation of shocks/turbulent acceleration sites in the acceleration region. From the best fit model we get that to respect the condition $\rho^{acc}_{p,B}>1.0$ we need to impose a lower limit of the ratio of relativistic electrons to cold protons  $N_e/N_p^{cold}<112$. This value is compatible with  the usual value of $N_e/N_p^{cold}=10$ \citep{Celotti2008} used in the case of relativistic jets with a leptonic radiative domination. 
    Moreover, we note that the value of $B_{acc}$ obtained from the best fit did not require a significant change in the value of $L_B$ as derived from the phenomenological model setup, and demonstrating that constraining $z_{acc}$ based on the value of $\nu_{t}$ is naturally in agreement with formation of mechanical compression in the jet when $U_p>U_B$.

    \item $m_{jet}=1.2$. The value of $m_{jet}$ is one the most critical, indeed it dictates the topology and intensity of the magnetic field beyond the acceleration region, and it is interesting to compare to the value of $m_B$ that is used to model the jet below the acceleration region. The initial guess based on the value of $\alpha_{R}$ has required  a small modification in order to reproduce the observed radio spectrum, and the final model naturally explains the almost flat radio spectrum as emission of the cooled electron leaving the acceleration region.
     %The increased value of $m_{jet}$ compared to $m_B$ might be related to rearrangement of the magnetic field as a consequence of the mechanic compression in the acceleration region. As a consequence, hinting for an increase in the expansion power of the jet in the radio region. Anyhow, 
  \end{itemize}

%%%%%%%%%%%%%%%%%%%%%%%%%%%%%%%%%%%%%%%%%%%%%%%%%%%%%%%%%%%%%%%%%%%%%%%%%%%%%%%%
%%%%%%%%%%%%%%%%%%%%%%%%%%%%%%%%%%%%%%%%%%%%%%%%%%%%%%%%%%%%%%%%%%%%%%%%%%%%%%%%

\section{Discussion and Conclusions}

As MAXI J\(1820+070\) was observed numerous times across the EM spectrum during its outburst, there are multiple works relevant to portions of our multi-wavelength analysis, though to date none study the source in such a complete picture as is presented with our model from \texttt{JetSeT}.  

%Our comparison will begins the irradiated disk component and then proceed to lower frequencies covering the jet components, respectively.

%As shown in Table~\ref{best-fit-table} and discussed in Section~\ref{xspec_jetset}, the \textt{JetSeT} analysis finds a disk temperature of \(\sim 0.13 \) keV compared to \(\sim 0.2 \) keV found by \cite{2019Natur.565..198K} during 6 observations with \textit{NICER} from 21 March (MJD 58201) until 8 May (MJD 58246) while \cite{2020arXiv200412946W} found slightly lower disk temperatures of \(\sim 0.18 \) keV from \textit{NICER} over the same time period.  However, over 8 \textit{NuSTAR} observations between 14 March (MJD 58191) and 28 June (MJD 58297),  \citep{2019MNRAS.490.1350B} report a temperature ranging from \( \sim 0.5 - 0.8 \) keV. 

\cite{2018ApJ...868...54S} provides the most direct comparison to the analysis in this work, though the source behavior is different before and after 26 March (MJD 58206).  They found that the optical and near-IR emission is not entirely from disk emission and thus included a power-law to their \texttt{diskir} model with a spectrum described by fixed parameters \(kT_{disk} = 0.35\) keV, \(L_C/L_D = 70\), \(f_{out} = 5 \times 10^{-3}\), and \(R_{out} = 10^5 R_{in}\).  Our fit found a considerably lower \(kT_{disk}\) (0.12 keV), \(L_C/L_D\) (4.7), and \(R_{out}\) (\(10^3 R_{in}\)), but a higher \(f_{out}\) (\(4 \times 10^{-2}\)). We note that the source behavior between the two observations is different with changes in the spectral hardness in hard X-rays \citep{2019ApJ...870...92R}, the development of type-C QPOs \citep{2020ApJ...889..142S}, and a reduction in the size of the corona \citep{2019Natur.565..198K} that can possibly explain the differences.  

Following the work of \cite{2019ApJ...879L...4M}, we explored the presence of disk wind signatures in our VLT/X-shooter optical spectrum, as this data-set falls between epochs 11 and 12 of \cite{2019ApJ...879L...4M} campaign and adds an epoch in their uncovered time window (between 26 March and 23 April). We focus our spectral analysis on the He\,I $\lambda$ 5876, He\,I $\lambda$ 6678 and H$\alpha$ wavelength regions. We found shallow p-cygni profiles and strong line asymmetries in all the three mentioned lines, while a broad outflow component is detected only in the red wing of the H$\alpha$. Among the observed absorption troughs, the one detected in the blue wing of He\,I $\lambda$ 5876 is the more prominent and it results in a terminal wind velocity $v_{t}$=880 km/s, which is consistent with the outflow velocity of $v\sim$900 km/s, derived from the H$\alpha$ redshifted broad component.
%The three lines are all characterised by strong asymmetries in their profiles, but they are all different from each other. While the \ion{He}{I} $\lambda$ 5876 shows a pronounced double-horned profile, the \ion{He}{I} $\lambda$ 6678 clearly shows the presence of three components, one is well centred on the rest-frame wavelength of the line, while the other two are blue-shifted and red-shifted of $\sim$450 km/s with respect the first component. Although the narrow H$\alpha$ also shows a double-peaked line profile, its shape is quite different from the \ion{He}{I} lines. In particular, it is well modelled by two narrow components, but, while the first and more prominent one is near the Ha rest-frame wavelength, the second is more faint and redshifted of $\sim$ 667 km/s.
These properties indicate that at this epoch optical disk winds are still present, although with slower velocities with respect to what found in \cite{2019ApJ...879L...4M}. The author also report an evolution of the line profiles during their monitoring campaign and our observation confirms this trend. In particular the observed H$\alpha$ profile can be interpreted as a continuation in the evolving pattern of the line between the epochs 9 and 12 shown in Figure 2 of \cite{2019ApJ...879L...4M}. Similar spectral variations were previously reported by \cite{2018ApJ...867L...9T} and ascribed to the orbital motion of the system.
Interestingly, some of the most conspicuous optical wind detections in \cite{2019ApJ...879L...4M} occur in epochs corresponding to the hard state of the source, when radio emission and strong jet activity are present \citep{2020NatAs.tmp....2B} and the peak of the optical outburst of the source is reported. This led the authors to the conclusion that the optical wind detected in MAXI J\(1820+070\) is simultaneous with the jet. Our wind signatures detection, together with the results from our broad band spectral analysis are consistent with this scenario.
%This led the authors to the conclusion that there is a significant jet contribution to the optical regime \cite[see also][]{2018ApJ...868...54S} and that the optical wind detected in MAXI J\(1820+070\) is simultaneous with the jet. Our wind signatures detection, together with the need for an additional radio component in our broad band spectral analysis are in line with this scenario. 

%\textcolor{orange}{+ RADIO JET COMPARISON WITH RUSSELL AND BRIGHT (GB) -- Combining observations from late March and early April, \cite{2018ATel11533....1R} performed an analysis of the compact jet emission and found spectral indices of \(\alpha_{thick} \sim 0.3\) and \(\alpha_{thin} \sim -0.7 \) for the optically thick and thin regions, respectively, when assuming a broken power-law model. Those authors suggested that the synchrotron peak may be located within the mid-IR to far-IR range. These results are comparable with the values found through our observations, from radio to optical bands when adopting a simple broken power-law model with $\alpha_{thick}=0.28\pm0.02$ and $\alpha_{thin}=-0.61\pm0.01$.}

Our phenomenological analysis of the compact jet found the data could be modelled by a broken power-law with $\alpha_{thick}=0.28\pm0.02$ and $\alpha_{thin}=-0.61\pm0.01$.  Combining observations from late March and early April, \cite{2018ATel11533....1R} performed a similar analysis and found spectral indices of \(\alpha_{thick} \sim 0.3\) and \(\alpha_{thin} \sim -0.7 \).  Building on \cite{2018ATel11533....1R}, \cite{2018ApJ...868...54S} estimated a transition frequency of $\sim 3\times 10^{13}$ Hz and a corresponding flux density of $\sim 0.4$ Jy.  From these values they determined \( B \sim 1 \times 10^4 \) G and \( R \sim 2 \times 10^9 \) cm using equations from \cite{2011PASJ...63S.785S}.  Our model peaks at  1.6$\pm$0.2$\times10^{13}$ Hz with a flux density of $\sim 0.35$ Jy thus resulting in similar values.

These values are in agreement with the phenomenological setup and with the best-fit model from \texttt{JetSeT}. In particular the \texttt{JetSeT} best-fit model gives a magnetic field in the acceleration region of $\approx 1.8 \times 10^4$ G, and a region radius of $\approx 2.6 \times 10^9$ cm.

The corresponding energy density of the magnetic field is $\approx 1.3\times10^7 \textrm{ erg/cm}^3$ compared to the values of \(8 \times 10^6 \textrm{erg/cm}^3 \) from \cite{2018ApJ...868...54S}.

Additionally, we identify a separate radio spectral components at frequencies below $\sim$10 GHz, showing an inverted power-law spectrum with slope $\alpha=0.11\pm0.02$. \cite{2020NatAs.tmp....2B}, collecting data from different epochs of VLA, Multi-Element Radio Linked Interferometer Network (eMERLIN), and Meer Karoo Array Telescope (MeerKAT) observations, could identify at least one ejected component during the transition from the hard to the soft state (mid-June to mid-September 2018). Though the source is unresolved  down to a sub-arcsec resolution in the VLA observations considered here (collected in a previous epoch), the presence of an additional low-frequency spectral component could suggest that the ejecta later detected by \cite{2020NatAs.tmp....2B} were already present at a sub-pc scale during the April 12th 2018 epoch considered here.

This component is represented  in the \jetset broadband model by the \texttt{RadioJet} component, and stems naturally from the cooling of the accelerated particle leaving the acceleration region.  Interestingly we find that the best-fit index $m_{jet}\approx 1.2$  predicts a radio spectral  index of $\alpha=1-1/m_{jet}\approx 0.166$ that is close to the value found in the  power-law fit. We note, that the small difference between the two values, is due to the fact the \jetset \texttt{RadioJet} model takes into account the data range from radio-to-mm frequencies,  differently from the  power-law fit, whose 
range extends up $\approx 10^{10}$ Hz.

In conclusion, our broadband analyses of MAXI J\(1820+070\) found the source in a hard state with parameters similar to what was reported by \cite{2018ApJ...868...54S} %while also exploring the behavior of a radio jet component using modelling in \texttt{JetSeT}.  
The \texttt{JetSeT} broadband model was  able to reproduce the full SED taking into account both the disk/corona emission, and the leptonic radiativelly dominated relativistic jet contribution. We found that the relativistic jet required a total energy of $L_{jet}\geq 8.0\times10^{37}$ erg/s, corresponding to 0.15 $L_{Edd}$. This value represents a lower limit, since we assume that the hadronic content of the jets is only in terms of cold protons, without a significant radiative contribution. The flat radio  spectral shape stems naturally from the synchroton cooling of the electrons in the acceleration regions, in agreement with previous analyses \citep{Kaiser2006,Peer:2009}.  In comparison, the accretion luminosity (\(6 \times 10^{37}\) erg/s) is comparable to the lower limit of the jet luminosity.  Thus in MAXI J\(1820+070\), it is possible for the jet to be powered predominately via accretion with only a small contribution from the Blanford-Znajek mechanism, which in this case cannot provide much power since the black hole spin is reported to be low \citep{Bassi2020,Zhao2020}.

\acknowledgments
We thank the Italian node of the European ALMA Regional Centre (ARC) for the support. 
JR and GB acknowledge financial support under the INTEGRAL ASI-INAF agreement 2019-35-HH.0 and ASI/INAF n. 2017-14-H.0. FO acknowledge the support of the H2020 European Hemera program, grant agreement No 730970.  The research leading to these results has received funding from the European Union’s Horizon 2020 Programme under the AHEAD2020 project (grant agreement n. 871158)
F.O. acknowledges the support of the H2020 European Hemera program, grant agreement No 730970, and the support of 
the GRAWITA/PRIN-MIUR project: "\textit{The new frontier of the Multi-Messenger Astrophysics: follow-up of electromagnetic transient counterparts of gravitational wave sources}".
Based on observations with INTEGRAL, an ESA project with instruments and science data centre funded by ESA member states (especially the PI countries: Denmark, France, Germany, Italy, Switzerland, Spain) and with the participation of Russia and the USA.
This research has made use of the services of the ESO Science Archive Facility. Based on observations collected at the European Southern Observatory under ESO programmes 2017.1.01103.T (ALMA) and 0101.D-0356(A) (VLT). ALMA is a partnership of ESO (representing its member states), NSF (USA), and NINS (Japan), together with NRC (Canada) and NSC and ASIAA (Taiwan), in co- operation with the Republic of Chile. The Joint ALMA Observatory is operated by ESO, AUI/NRAO, and NAOJ. 
The National Radio Astronomy Observatory is a facility of the National Science Foundation operated under cooperative agreement by Associated Universities, Inc.
%
%%%%%%%%%%%%%%%%%%%%%%%%%%%%%%%%%%%%%%%%%%%%%%%%%%%%%%%%%%%%%%%%%%%%%%%%%%%%%%%%
%%%%%%%%%%%%%%%%%%%%%%%%%%%%%%%%%%%%%%%%%%%%%%%%%%%%%%%%%%%%%%%%%%%%%%%%%%%%%%%%

\bibliography{XRB}{}
\bibliographystyle{aasjournal}

%%%%%%%%%%%%%%%%%%%%%%%%%%%%%%%%%%%%%%%%%%%%%%%%%%%%%%%%%%%%%%%%%%%%%%%%%%%%%%%%
\end{document}